# Giant magnetic exchange coupling in rhombus-shaped nanographenes with zigzag periphery

Shantanu Mishra[1], Xuelin Yao[2], Qiang Chen[2], Kristjan Eimre[1], Oliver Gröning[1], Ricardo Ortiz[3,4], Marco Di Giovannantonio[1], Juan Carlos Sancho-García[4], Joaquín Fernández-Rossier[3,5], Carlo A. Pignedoli[1], Klaus Müllen[2], Pascal Ruffieux[1], Akimitsu Narita[2,6]*, Roman Fasel[1,7]*

[1]nanotech@surfaces Laboratory, Empa—Swiss Federal Laboratories for Materials Science and Technology, Dübendorf, Switzerland

[2]Department of Synthetic Chemistry, Max Planck Institute for Polymer Research, Mainz, Germany

[3]Department of Applied Physics, University of Alicante, Sant Vicent del Raspeig, Spain

[4]Department of Chemical Physics, University of Alicante, Sant Vicent del Raspeig, Spain

[5]QuantaLab, International Iberian Nanotechnology Laboratory, Braga, Portugal

[6]Organic and Carbon Nanomaterials Unit, Okinawa Institute of Science and Technology Graduate University, Okinawa, Japan

[7]Department of Chemistry and Biochemistry, University of Bern, Bern, Switzerland

*Corresponding authors: narita@mpip-mainz.mpg.de and roman.fasel@empa.ch

**Nanographenes with zigzag edges are predicted to manifest non-trivial π-magnetism resulting from the interplay of hybridization of localized frontier states and Coulomb repulsion between valence electrons. This provides a chemically tunable platform to explore quantum magnetism at the nanoscale and opens avenues toward organic spintronics. The magnetic stability in nanographenes is thus far limited by the weak magnetic exchange coupling which remains below the room temperature thermal energy. Here, we report the synthesis of large rhombus-shaped nanographenes with zigzag periphery on gold and copper surfaces. Single-molecule scanning probe measurements unveil an emergent magnetic spin-singlet ground state with increasing nanographene size. The magnetic exchange coupling in the largest nanographene, determined by inelastic electron tunneling spectroscopy, exceeds 100 meV or 1160 K, which outclasses most inorganic nanomaterials and remarkably survives on a metal electrode.**

Magnetism in solids is usually associated to *d*- or *f*-block elements. However, since the isolation of graphene, the field of carbon magnetism has gained increased traction[1]. Though ideal graphene is a diamagnetic semimetal, many of its derivative nanostructures (nanographenes) are predicted to manifest magnetism which is distinct from magnetism in molecules and solids containing transition metal atoms. First, in contrast to the localized nature of magnetic moments in transition metal atoms, unpaired electrons in nanographenes are hosted by molecular π-orbitals, which extend over several carbon atoms. Second, emergence of magnetic moments in nanographenes results from two competing phenomena: (1) hybridization energy, which is responsible for the formation of the highest occupied and lowest unoccupied molecular orbitals (HOMO and LUMO), leading to a non-magnetic (closed-shell) ground state, and (2) electrostatic Coulomb repulsion between valence electrons, which promotes the formation of singly occupied molecular orbitals (SOMOs) hosting unpaired spins, resulting in a magnetic (open-shell) ground state. These two energy scales, along with the magnetic ordering between unpaired spins, can be efficiently tailored through a rational design of the shape, size and edge structure of nanographenes. It has long been known that hybridization energy may be strongly reduced or even completely removed in nanographenes with zigzag edges[2–4], with higher anthenes[5,6] and zethrenes[7,8] being typical examples of nanographenes with open-shell characters. In recent years, this concept has also been utilized to fabricate open-shell nanographenes on surfaces, whose magnetic ground states have been directly evidenced through detection of Kondo interactions between localized spins and conduction electrons of metal surfaces, and spin excitations of coupled spin systems with the scanning tunneling microscope (STM)[9–11].

Experimental realization of nanographenes containing zigzag edges is largely restricted to structures with a mixture of zigzag and armchair edges, with prominent examples being anthenes, zethrenes and periacenes[12–14], and synthesis of nanographenes with all sides consisting of zigzag edges (zigzag nanographenes, ZNGs) has proven challenging. Triangular ZNGs[15], which exist as neutral radicals[16,17], have been recently synthesized on metal and insulator surfaces under ultrahigh vacuum[18–20]. However, evidence of magnetism in these nanographenes is indirect and relies on the detection of SOMOs by scanning tunneling spectroscopy (STS). On the other hand, solution synthesis of large ZNGs is limited to closed-shell systems[21–24]. We report here the on-surface synthesis of rhombus-shaped ZNGs—hereafter [$n$]-rhombenes, where $n$ is the number of benzenoid rings along an edge, with $n = 4$ ($C_{48}H_{18}$, **1**) and 5 ($C_{70}H_{22}$, **2**)—which represent the largest ZNGs synthesized to date (Fig. 1). Our inelastic electron tunneling spectroscopy (IETS) measurements, supported by theoretical models, show that [$n$]-rhombenes acquire an open-shell ground state with increasing size. The magnetic exchange coupling (MEC) in the largest nanographene, **2**, is directly determined to be 102 meV, which exceeds the room temperature thermal energy by a factor of four, and surpasses the MEC in most known transition metal nanomagnets[25].

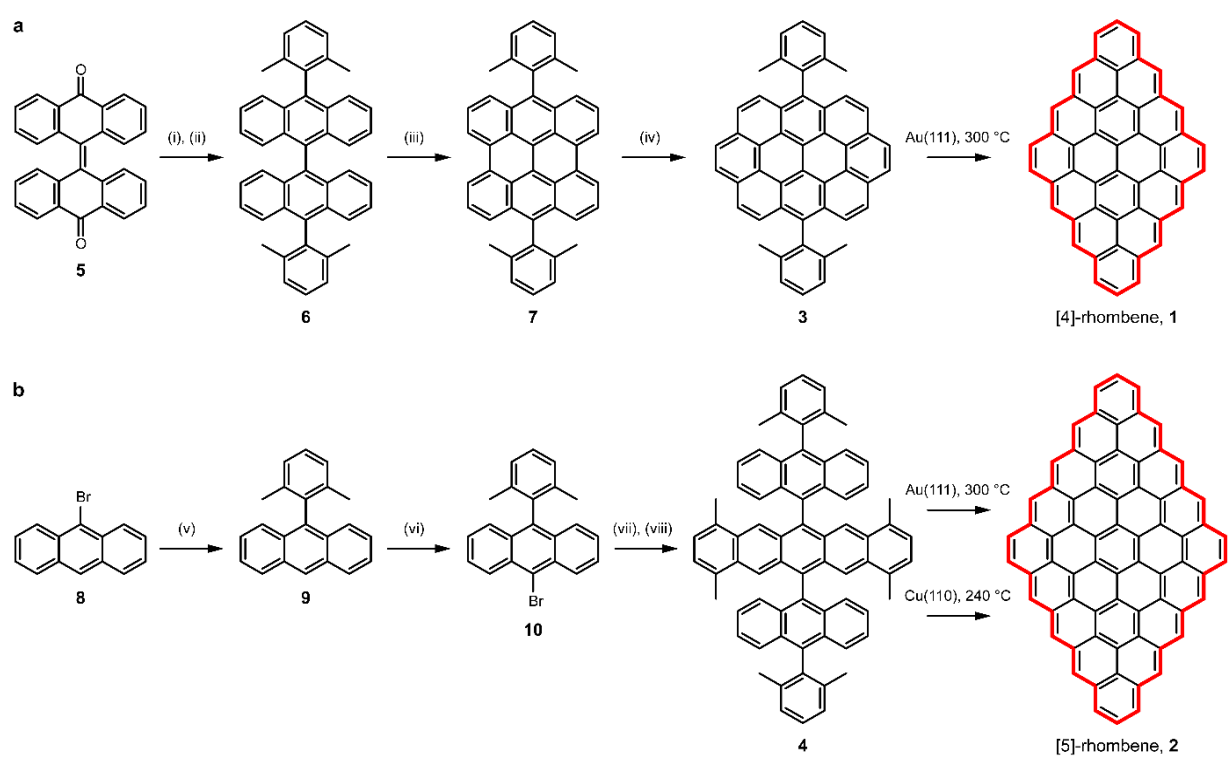

**Figure 1 | Synthesis of [4]- and [5]-rhombenes. a,b**, Combined in-solution and on-surface synthetic routes toward [4]- (**a**) and [5]-rhombenes (**b**). The zigzag peripheries of **1** and **2** are highlighted in red. Conditions: (i) 2,6-dimethylphenylmagnesium bromide, THF, r.t., overnight; thereafter $CH_3COOH$, r.t., 10 min; (ii) NaI, $NaH_2PO_2 \cdot H_2O$, $CH_3COOH$, reflux, 2 h, 52% yield in two steps; (iii) DDQ, $Sc(OTf)_3$, $CF_3SO_3H$, chlorobenzene, 140 °C, 18 h, 25% yield; (iv) 2-nitroethanol, phthalic anhydride, $o$-xylene, 165 °C, 24 h, 68% yield; (v) 2,6-dimethylphenyl-boronic acid, $Pd(OAc)_2$, SPhos, $K_3PO_4$, toluene, 120 °C, 18 h, 73% yield; (vi) Bromine, $CCl_4$, r.t., 15 min, 96% yield; (vii) $n$-BuLi, diethyl ether, −78 °C, 1 h, 7,10-dimethyltetracene-5,12-dione, 0 °C, 48 h; (viii) NaI, $NaH_2PO_2 \cdot H_2O$, $CH_3COOH$, reflux, 6 h, 29% yield in two steps.

## Results and discussion

Our synthetic strategy toward [$n$]-rhombenes involves the design of molecular precursors 7,14-bis(2,6-dimethylphenyl)ovalene (**3**) and 6,13-bis{10-(2,6-dimethylphenyl)anthracen-9-yl}-1,4,8,11-tetramethylpentacene

(**4**) (Fig. 1), which are expected to undergo surface-catalyzed cyclodehydrogenation and oxidative cyclization of methyl groups, thereby yielding **1** and **2**, respectively. The synthesis of **3** (Fig. 1a) was performed starting from bisanthrone (**5**), which was treated with (2,6-dimethylphenyl)magnesium bromide followed by dehydroxylation under an acidic condition to provide 10,10'-bis(2,6-dimethylphenyl)-9,9'-bianthracene (**6**). The oxidative cyclodehydrogenation of **6** gave 7,14-bis(2,6-dimethylphenyl)bisanthene (**7**), which was subjected to two-fold Diels-Alder addition with nitroethylene to obtain **3**. The synthesis of **4** (Fig. 1b) was performed through the Suzuki coupling of 9-bromoanthracene (**8**) with (2,6-dimethylphenyl)boronic acid to afford 9-(2,6-dimethylphenyl)anthracene (**9**), followed by its bromination to yield 9-bromo-10-(2,6-dimethylphenyl)anthracene (**10**). Subsequently, **10** was lithiated to {10-(2,6-dimethylphenyl)anthracen-9-yl}lithium and reacted with 7,10-dimethyltetracene-5,12-dione to provide **4** after reduction. Toward the synthesis of **1**, **3** was deposited on a Au(111) surface and annealed to 300 °C to promote the on-surface reactions. High-resolution STM imaging elucidated that 92% of the molecules on the surface exhibit a uniform rhomboid shape (Fig. 2a,b), and chemical structure determination via ultrahigh-resolution STM imaging[26,27] unambiguously proved the formation of **1** (Fig. 2c). We further attempted the on-surface synthesis of **2** from **4** using a similar strategy. The overview STM image after annealing **4** at 300 °C on Au(111) revealed the predominance of covalently coupled molecular clusters (Fig. 2d and Supplementary Fig. 1), and in contrast to **1**, we rarely found isolated molecules on the surface. Figure 2e,f show the high-resolution and ultrahigh-resolution STM images of an isolated molecule, respectively, demonstrating the successful formation of **2**. The pronounced intermolecular reactions of **2** is indicative of a considerably higher reactivity of **2** compared to **1**, as the propensity for **1** under identical synthetic conditions is to remain isolated on the surface. As we demonstrate below, this drastic difference in the reactivity of **1** and **2** is due to a larger zigzag periphery in **2**, leading to an open-shell ground state.

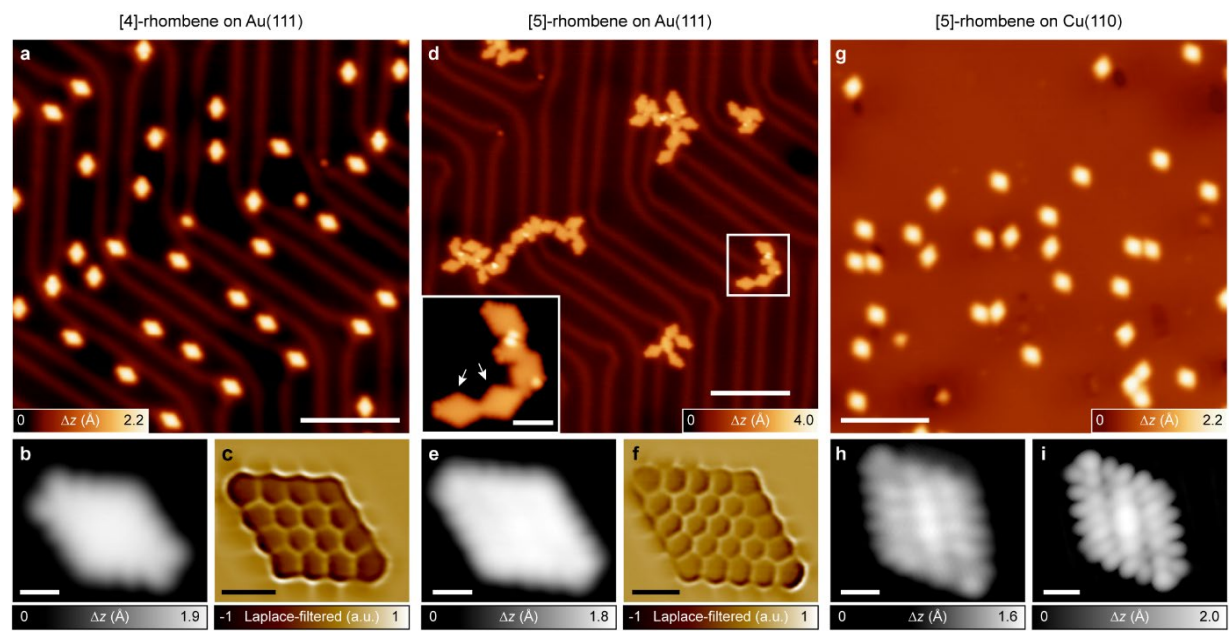

**Figure 2 | On-surface synthesis and STM characterization of [4]- and [5]-rhombenes. a**, Overview STM image after annealing **3** on Au(111) at 300 °C ($V = -300$ mV, $I = 120$ pA). **b,c**, High-resolution ($V = -330$ mV, $I = 50$ pA) (**b**) and Laplace-filtered ultrahigh-resolution ($V = -5$ mV, $I = 50$ pA, $\Delta z = -0.8$ Å) (**c**) STM images of **1** on Au(111). a.u. denotes arbitrary units. **d**, Overview STM image after annealing **4** on Au(111) at 300 °C ($V = -400$ mV, $I = 100$ pA). Inset: High-resolution STM image of a cluster, where two constituent molecules corresponding to **2** are marked with arrows. **e,f**, High-resolution ($V = -10$ mV, $I = 50$ pA) (**e**) and Laplace-filtered ultrahigh-resolution ($V = -5$ mV, $I = 50$ pA, $\Delta z = -0.9$ Å) (**f**) STM images of **2** on Au(111). **g**, Overview STM image after annealing **4** on Cu(110) at 240 °C ($V = -200$ mV, $I = 100$ pA). **h,i**, High-resolution ($V = -100$ mV, $I = 50$ pA) (**h**) and DFT-simulated ($V = -100$ mV) (**i**) STM images of **2** on Cu(110). Scale bars: 10 nm (**a,d,g**), 2 nm (inset **d**) and 0.5 nm (all other panels). Images in **b,e,h** were acquired with carbon monoxide (CO)-functionalized tips.

To circumvent the problem of the limited yield of **2** on Au(111), we conducted the synthesis of **2** on Cu(110) surface. In contrast to densely-packed (111) surfaces, which exhibit weak interactions with molecular adsorbates, the more open (110) surfaces feature comparatively stronger hybridization between molecular orbitals and the metal $d$-states due to energetic upshift of the $d$-band center[28]. As a result, molecular mobility may be substantially reduced[29] which could prevent intermolecular reactions and increase the yield of isolated molecules. Figure 2g presents an overview STM image after deposition of **4** on Cu(110) and annealing to 240 °C. In contrast to Au(111), the reaction products on Cu(110) nearly exclusively consist of isolated molecules exhibiting a rhomboid shape. The excellent match between the experimental STM image of the molecules (Fig. 2h) and the density functional theory (DFT)-simulated STM image of **2** on Cu(110) (Fig. 2i) proves the successful formation of **2**.

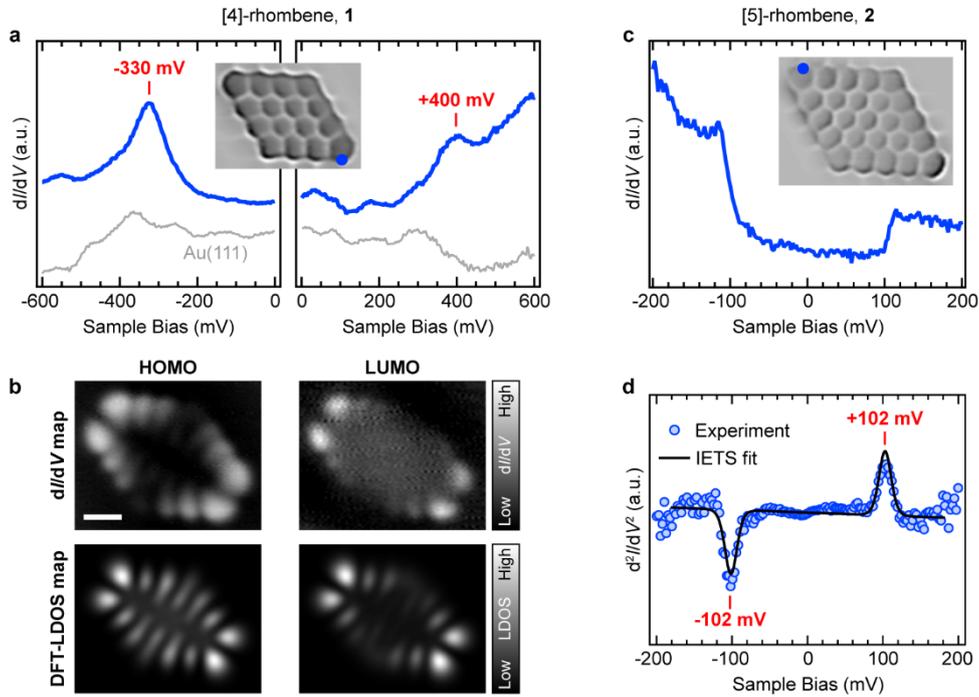

**Figure 3 | Electronic and magnetic characterization of [4]- and [5]-rhombenes. a**, d$I$/d$V$ spectra acquired on **1** with a CO-functionalized tip revealing HOMO and LUMO resonances at −330 mV and +400 mV, respectively (open feedback parameters: $V$ = −600 mV, $I$ = 200 pA (left panel) and $V$ = +600 mV, $I$ = 500 pA (right panel); $V_{rms}$ = 16 mV). **b**, Constant-current d$I$/d$V$ maps acquired with a CO-functionalized tip (upper panels) and corresponding DFT-LDOS maps (lower panels) at the HOMO and LUMO resonances of **1** ($V$ = −330 mV, $I$ = 150 pA; $V_{rms}$ = 20 mV (HOMO d$I$/d$V$ map) and $V$ = +380 mV, $I$ = 160 pA; $V_{rms}$ = 24 mV (LUMO d$I$/d$V$ map)). **c,d**, Background-subtracted d$I$/d$V$ spectrum acquired on **2** revealing inelastic excitation steps (open feedback parameters: $V$ = −200 mV, $I$ = 200 pA; $V_{rms}$ = 4 mV) (**c**); and corresponding IETS spectrum (filled circles, open feedback parameters: $V$ = −200 mV, $I$ = 2.8 nA; $V_{rms}$ = 10 mV), with fit to the experimental data using the Heisenberg dimer model (solid line) (**d**). The spin excitation threshold is extracted to be ±102 mV. Acquisition positions for the respective spectra are marked with a filled circle in the insets of **a,c**. The data in **c,d** are acquired on different molecules with different tips. Scale bar: 0.5 nm.

We employ STS to experimentally probe the electronic structures of **1** and **2** on Au(111). Differential conductance spectroscopy (d$I$/d$V$, where $I$ and $V$ denote current and voltage, respectively) on **1** reveals peaks in the local density of states (LDOS) at −330 mV and +400 mV (Fig. 3a). Spatially resolved d$I$/d$V$ maps at the respective energies exhibit excellent agreement with the DFT-LDOS maps of the HOMO and LUMO of **1** (Fig. 3b and Supplementary Fig. 2), confirming the spectroscopic features to be molecular orbital resonances, with the HOMO-LUMO gap being 730 meV. In contrast, d$I$/d$V$ spectroscopy on **2** reveals abrupt stepwise change in conductance symmetric around the Fermi energy, indicative of an inelastic excitation[30] (Fig. 3c). The excitation threshold, extracted from a fit to the corresponding IETS spectrum[31], equals ±102 meV (Fig. 3d).

To investigate if the inelastic excitation may be ascribed to a spin excitation[25], we performed DFT calculations to unravel the magnetic ground states of [$n$]-rhombenes. Since [$n$]-rhombenes have an equal number of carbon atoms in the two interpenetrating triangular sublattices of their honeycomb lattice, the ground state, as per Lieb's theorem for bipartite lattices[32,33], is expected to be a singlet, that is, the total spin quantum number $S$ = 0. However, this rule does not predict whether a system with $S$ = 0 ground state is an open-shell or a closed-shell singlet. This complementary information is obtained from spin-polarized DFT calculations. Figure 4a presents the energetic difference between the open-shell singlet and closed-shell states for a series of [*n*]-rhombenes, showing a size-dependent onset of magnetism. While for $n \leq 4$, the ground state is closed-shell, an open-shell singlet ground state emerges for $n \geq 5$, in agreement with previous reports[34,35] and in support of our experimental observations. In particular, for **2**, the open-shell triplet ($S = 1$) and closed-shell states are 100 meV and 122 meV higher in energy, respectively, compared to the open-shell singlet ground state.

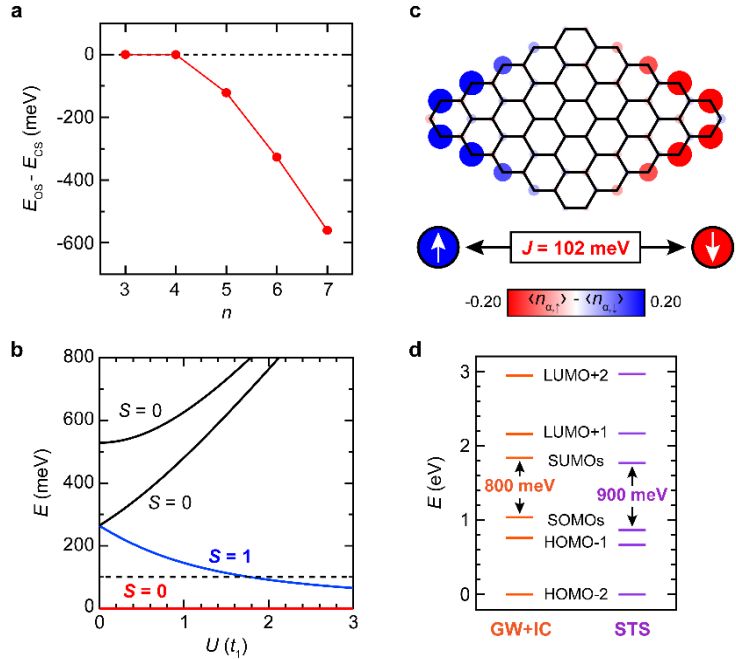

**Figure 4 | Theoretical calculations. a**, DFT-computed energy difference between the open-shell singlet (OS) and closed-shell (CS) states of [*n*]-rhombenes, for $n = 3\text{-}7$. **b**, Excitation spectrum of **2** as a function of $U$ ($U$ is scaled with respect to $t_1$), obtained with an exact diagonalization of the Hubbard model. The energies are given with respect to the $S = 0$ open-shell singlet ground state (red curve). The first excited state corresponds to an open-shell triplet (blue curve). The dashed line indicates the experimental singlet-triplet gap of 102 meV obtained from IETS measurements. **c**, Mean-field Hubbard spin polarization plot of the open-shell singlet ground state of **2** calculated at $U = t_1$, with schematic illustration of the exchange coupling of spins in **2** ($J$ denotes the MEC). Blue and red isosurfaces denote spin up and spin down populations. **d**, Energies of the HOMO–2 to LUMO+2 resonances of **2** on Au(111) from GW+IC calculations (orange markers) and STS experiments (purple markers), with the Coulomb splittings indicated in bold. Energies of the calculated and experimental levels are aligned at HOMO–2.

Furthermore, we calculated the magnetic excitation spectrum of **2** via an exact diagonalization of the Hubbard model at half filling[36,37]—known to give results in agreement with advanced quantum chemistry methods[37]—in the complete active space approximation with two electrons and two orbitals, with third nearest neighbor hopping. Figure 4b presents the energies of the manifold of excited states of **2** relative to its open-shell singlet ground state, as a function of the on-site Coulomb repulsion $U$. The first excited state is an open-shell triplet, with the corresponding singlet-triplet gap in the range of 75–150 meV for $t_1 \leq U \leq 2.5 t_1$ (where $t_1 = 2.7$ eV is the nearest-neighbor hopping parameter), in close correspondence to the experimental excitation threshold of 102 meV. Figure 4c presents the mean-field Hubbard spin polarization plot of **2**, where spin up and spin down populations are sublattice-polarized and localized at the opposite ends of the molecule. Therefore, while $S = 0$, a local spin polarization is maintained at each end. The corresponding picture in the case of a closed-shell ground state would entail equal population of spin up and spin down electrons at every carbon atom of **2**.

Finally, to establish a quantitative link between the open-shell character of **2** on Au(111) to experimental measures of electronic structure (that is, orbital resonances), we performed many-body perturbation theory GW calculations[38] (where G and W denote Green's function and screened Coulomb potential, respectively), including screening effects from the underlying surface[39] (that is, GW+IC, where IC denotes image charge). Figure 4d pre-sents a comparative energy spectrum of the HOMO–2 to LUMO+2 resonances of **2** determined from STS (Supplementary Figs. 3 and 4) and GW+IC calculations for the open-shell singlet state of **2**. Both the Coulomb gap and the relative energies of the HOMO–2 to LUMO+2 resonances obtained from GW+IC calculations exhibit a remarkable agreement with the corresponding experimental values, thus providing a striking confirmation of the open-shell singlet ground state of **2** on Au(111). We also conducted d$I$/d$V$ spectroscopy of **2** on Ag(111) to probe any change in the MEC (Supplementary Fig. 5). We detect a charge transfer from Ag(111) to **2**, leading to complete filling of the singly unoccupied molecular orbitals (SUMOs) and therefore a closed-shell ground state of **2**. In accordance with this observation, no spin excitations are seen for **2** on Ag(111).

For robust spin-logic operations at practical temperatures, it is imperative that the MEC exceeds the Landauer limit of minimum energy dissipation[40] at room temperature[41,42], $k_B T \ln(2) \approx 18$ meV, where $k_B$ is the Boltzmann's constant and $T$ is the temperature. The weak MEC of few millielectronvolts commonly found in transition metal nanomagnets limits operations of devices based on such materials to cryogenic temperatures. In contrast, the experimental singlet-triplet gap of 102 meV of **2**, which provides a direct measure of the MEC, is more than five times larger than the Landauer limit—highly promising for room temperature-stable spintronics. On a fundamental note, our results demonstrate that the synthesis of ZNGs with controlled size and shape allows building nanostructures with robust all-carbon magnetism, which mimic elementary building blocks to explore quantum magnetism. Given the robust MEC and negligible magnetic anisotropy in [$n$]-rhombenes, construction of their nanoscale lattices through established on-surface synthetic techniques could pave the way for exploration of exotic low-dimensional quantum phases of matter[43,44] in purely organic systems.

**Methods**

**Synthesis of molecular precursors.** The detailed solution synthesis of molecular precursors **3** and **4**, and associated characterization data are reported in Supplementary Schemes 1 and 2, and Supplementary Figs. 6–21.

**Sample preparation and STM/STS measurements.** STM measurements were performed in a commercial low-temperature STM from Scienta Omicron operating at a temperature of 4.5 K and base pressure below 5×10$^{-11}$ mbar. Au(111), Ag(111) and Cu(110) single crystal surfaces were prepared by Ar$^+$ sputtering and annealing cycles. Powder samples of precursors **3** and **4** were contained in quartz crucibles and sublimed from a home-built evaporator at 270 °C and 310 °C, respectively, onto single crystal surfaces held at room temperature. STM images and d$I$/d$V$ maps were recorded in constant-current (that is, closed feedback loop) mode, and d$I$/d$V$ and d$^2I$/d$V^2$ spectra were recorded in constant-height (that is, open feedback loop) mode. d$I$/d$V$ and d$^2I$/d$V^2$ measurements were obtained with a lock-in amplifier operating at a frequency of 860 Hz. Modulation voltages for each measurement are reported as root mean squared amplitude ($V_{rms}$). Tunneling bias voltages are provided with respect to the sample. Unless otherwise noted, STM and STS measurements were performed with metallic tips. Ultrahigh-resolution STM images were acquired by scanning the molecules with CO-functionalized tips in constant-height mode, and the current channel is displayed. For ultrahigh-resolution STM images, $\Delta z$ indicates lowering of the tip height after opening the feedback loop at the center of the molecules. CO molecules were deposited on a cold sample (with a maximum sample temperature of 13 K) containing reaction products and post-deposited NaCl islands, which facilitate CO identification and pick up. The data reported in this study were processed with WaveMetrics Igor Pro or WSxM[45] software.

**Tight-binding calculations.** Tight-binding calculations for obtaining spin polarization plots and LDOS maps of **2** have been performed by numerically solving the mean-field Hubbard Hamiltonian with third nearest neighbor hopping

$$\hat{H}_{MFH} = \sum_j \sum_{\langle \alpha,\beta \rangle_j, \sigma} -t_j c^\dagger_{\alpha,\sigma} c_{\beta,\sigma} + U \sum_{\alpha,\sigma} \langle n_{\alpha,\sigma} \rangle n_{\alpha,\bar{\sigma}} - U \sum_\alpha \langle n_{\alpha,\uparrow} \rangle \langle n_{\alpha,\downarrow} \rangle. \tag{1}$$

Here, $c^\dagger_{\alpha,\sigma}$ and $c_{\beta,\sigma}$ denote the spin selective ($\sigma \in \{\uparrow,\downarrow\}$ with $\bar{\sigma} \in \{\downarrow,\uparrow\}$) creation and annihilation operator at sites $\alpha$ and $\beta$, $\langle\alpha,\beta\rangle_j$ ($j = \{1,2,3\}$) denotes the nearest neighbor, second nearest neighbor and third nearest neighbor sites for $j = 1, 2$ and $3$, respectively, $t_j$ denotes the corresponding hopping parameters (with $t_1 = 2.7$ eV, $t_2 = 0.1$ eV and $t_3 = 0.4$ eV for nearest neighbor, second nearest neighbor and third nearest neighbor hopping[46]), $U$ denotes the on-site Coulomb repulsion, $n_{\alpha,\sigma}$ denotes the number operator, and $\langle n_{\alpha,\sigma}\rangle$ denotes the mean occupation number at site $\alpha$. Orbital electron densities, $\rho$, of the $n^{\text{th}}$-eigenstate with energy $E_n$ have been simulated from the corresponding state vector $a_{n,i,\sigma}$ by

$$\rho_{n,\sigma}(\vec{r}) = \left|\sum_i a_{n,i,\sigma}\phi_{2p_z}(\vec{r}-\vec{r}_i)\right|^2, \qquad (2)$$

where $i$ denotes the atomic site index and $\phi_{2p_z}$ denotes the Slater $2p_z$ orbital for carbon.

**Exact diagonalization of the Hubbard Hamiltonian.** The exact diagonalization of the Hubbard Hamiltonian is carried out in the two-level system at half filling[47], for which the energy levels $\pm U_M/2$ and $\pm\sqrt{4t_M^2 + U_M^2/4}$ of the system can be found analytically. The two levels, which are taken into consideration, are the sublattice-polarized zero energy states: $|1\rangle = 1/\sqrt{2}(|HOMO\rangle + |LUMO\rangle)$ and $|2\rangle = 1/\sqrt{2}(|HOMO\rangle - |LUMO\rangle)$, such that $t_M = \langle 1|\hat{H}|2\rangle$ and $U_M = U\sum_i|\langle i|1\rangle|^4$. Here, $\hat{H}$ denotes the third nearest neighbor tight-binding Hamiltonian, $U$ the on-site Coulomb repulsion of the carbon $2p_z$ orbital and $|i\rangle$ the $2p_z$ orbital of atomic site $i$.

**DFT and GW calculations.** The equilibrium geometries of the molecules adsorbed on Au(111) and Cu(110) surfaces were obtained with the CP2K code[48] implementing DFT within a mixed Gaussian plane waves approach[49]. The surface/adsorbate systems were modeled within the repeated slab scheme[50] in the following manner: (1) for molecules on Au(111), the simulation cell contained 4 atomic layers of Au along the [111] direction and a layer of hydrogen atoms to passivate one side of the slab in order to suppress one of the two Au(111) surface states, and (2) for molecules on Cu(110), the simulation cell contained 8 atomic layers of Cu along the [110] direction. 40 Å of vacuum was included in the simulation cell to decouple the system from its periodic replicas in the direction perpendicular to the surface. The electronic states were expanded with a TZV2P Gaussian basis set for C and H[51] and a DZVP basis set for Au and Cu. A cutoff of 600 Ry was used for the plane wave basis set. Norm-conserving Goedecker-Teter-Hutter pseudopotentials[52] were used to represent the frozen core electrons of the atoms. We used the PBE parameterization for the generalized gradient approximation of the exchange correlation functional[53]. We used the D3 scheme proposed by Grimme et al. to account for van der Waals interactions[54]. The Au surface was modeled by a supercell of 41.27×40.85 Å$^2$ corresponding to 224 surface units, and the Cu surface was modeled by a supercell of 43.20×35.65 Å$^2$ corresponding to 168 surface units. To obtain the equilibrium geometries, we kept the atomic positions of the bottom two layers of the Au slab and the bottom four layers of the Cu slab fixed to the ideal bulk positions, and all other atoms were relaxed until forces were lower than 0.005 eV/Å. Simulated STM images[55,56] within the Tersoff-Hamann approximation[57,58] were obtained by extrapolating the electronic orbitals to the vacuum region in order to correct the wrong decay of charge density in vacuum due to the localized basis set[58].

The gas phase geometry optimization and energy calculations for the restricted and unrestricted DFT were performed with the Martyna-Tuckerman Poisson solver together with a cell of size equal to double of the molecular bounding box plus 8 Å, while other inputs were kept equivalent to the slab calculation.

CP2K code was also used to perform the eigenvalue-self consistent GW calculations on the isolated molecular geometry corresponding to the adsorption conformation. The calculation was performed based on the unrestricted DFT-PBE wave functions. We employed the GTH pseudopotentials and analytic continuation with a two-pole model. The aug-DZVP basis set from Wilhelm et al. was used[59]. To account for screening by the underlying metal surface, we applied the image charge model[39]. To determine the image plane position with respect to the molecular geometry, we used a distance of 1.42 Å between the image plane and the first surface layer, as reported by Kharche and Meunier[60]. The calculations were performed via workflows based on the AiiDA platform[61].

**Data availability.** The crystallographic data have been deposited at the Cambridge Crystallographic Data Centre (CCDC) under CCDC numbers 1978171 (**3**) and 1978172 (**4**), and copies can be obtained free of charge from www.ccdc.cam.ac.uk/data_request/cif. Other data supporting the findings of this study are available from the corresponding authors on reasonable request.

**Code availability.** The tight-binding calculations were performed using a custom-made code on the WaveMetrics Igor Pro platform. Details of the tight-binding code can be obtained from the corresponding authors on reasonable request.

## Acknowledgements

We thank D. Passerone for discussions and D. Schollmeyer for single-crystal X-ray analysis. This work was supported by the Swiss National Science Foundation (grant numbers 200020-182015 and IZLCZ2-170184), the NCCR MARVEL funded by the Swiss National Science Foundation (grant number 51NF40-182892), the European Union's Horizon 2020 research and innovation program (grant number 785219, Graphene Flagship Core 2), the Office of Naval Research (N00014-18-1-2708), the Max Planck Society, Generalitat Valenciana and Fondo Social Europeo (grant number ACIF/2018/175), MINECO-Spain (grant number MAT2016-78625), and Portuguese FCT (grant number UTAPEXPL/ NTec/0046/2017). Computational support from the Swiss Supercomputing Center (CSCS) under project ID s904 is gratefully acknowledged.

## Author Contributions

R.F., P.R., A.N. and K.M. conceived the project. Q.C. and X.Y. synthesized and characterized the precursor molecules. S.M. performed the STM experiments. S.M. analyzed the data with contributions from M.D. K.E. and C.A.P. performed the DFT and GW calculations. R.O., J.F.R., J.C.S.G., O.G. and S.M. performed the tight-binding and Hubbard calculations. All authors contributed to discussing the results and writing the manuscript.

## Competing Interests

The authors declare no competing interests.

# Supplementary Information

# Giant magnetic exchange coupling in rhombus-shaped nanographenes with zigzag periphery

Shantanu Mishra, Xuelin Yao, Qiang Chen, Kristjan Eimre, Oliver Gröning, Ricardo Ortiz, Marco Di Giovannantonio, Juan Carlos Sancho-García, Joaquín Fernández-Rossier, Carlo A. Pignedoli, Klaus Müllen, Pascal Ruffieux, Akimitsu Narita, Roman Fasel

## CONTENTS

# 1. Additional STM/STS data and theoretical calculations.

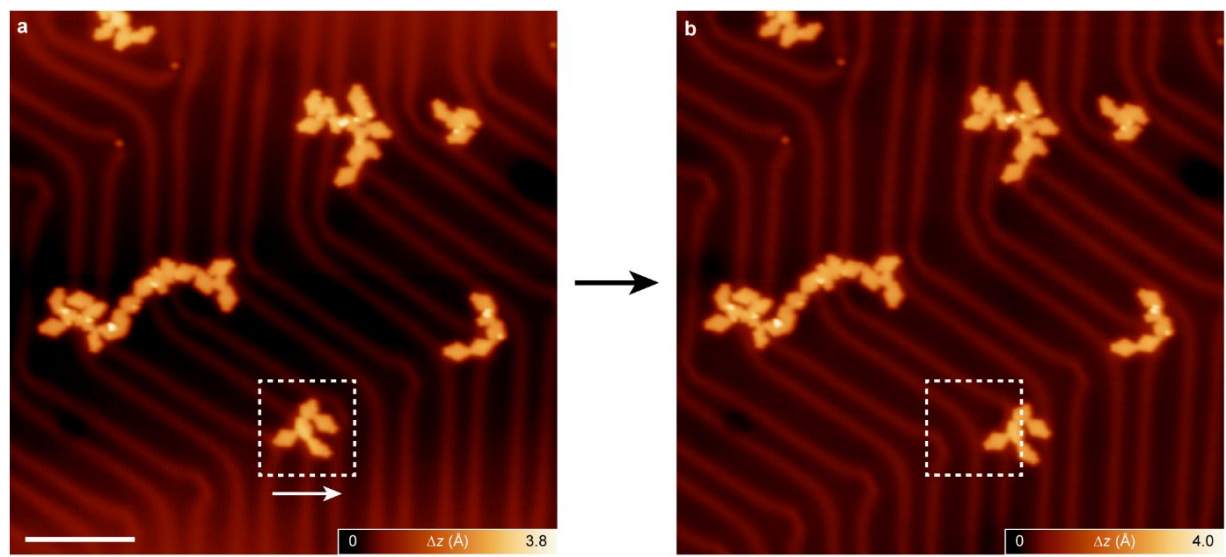

**Supplementary Figure 1 | Lateral manipulation of a molecular cluster. a**, Overview STM image after annealing **4** on Au(111) at 300 °C. The cluster highlighted with a dashed square was chosen for manipulation. The tip trajectory for manipulation is highlighted with an arrow. **b**, STM image of the same area after the manipulation event (shown in Fig. 2d). The dashed square corresponds to the same location as in **a**, and highlights considerable displacement of the cluster from its original location. The intact displacement of the cluster as a whole evidences that the constituent molecules are covalently coupled. Scanning parameters: $V = -400$ mV, $I = 100$ pA. Manipulation parameters: $V = -10$ mV, $I = 5$ nA. Scale bar: 10 nm.

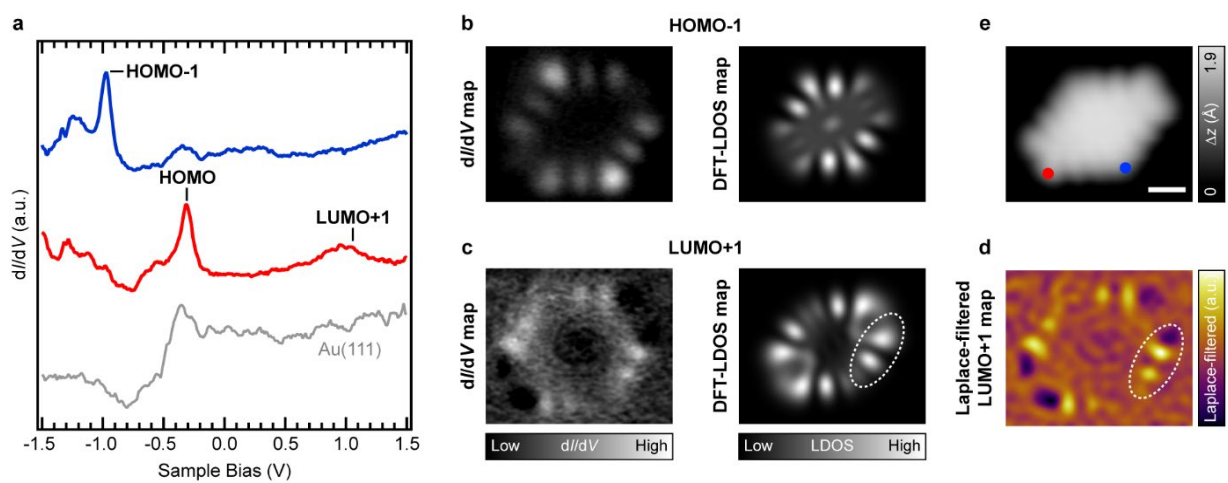

**Supplementary Figure 2 | Detection of non-frontier molecular orbitals of 1 on Au(111). a**, Long-range d$I$/d$V$ spectra acquired on **1** with a CO-functionalized tip. Acquisition positions are marked with filled circles in **e**. Apart from the HOMO resonance at −330 mV, a prominent HOMO−1 resonance is found at −970 mV, along with a weak and broad LUMO+1 resonance at ~ +1.1 V. The weak LUMO resonance at +400 mV is not visible here. Open feedback parameters: $V = -1.5$ V, $I = 200$ pA; $V_{rms} = 20$ mV. **b**, Constant-current d$I$/d$V$ map at the HOMO−1 resonance of **1** (left panel; $V = -970$ mV, $I = 150$ pA; $V_{rms} = 24$ mV) and the corresponding HOMO−1 DFT-LDOS map (right panel). **c**, Constant-current d$I$/d$V$ map at the LUMO+1 resonance of **1** acquired with a CO-functionalized tip (left panel; $V = +1.3$ V, $I = 150$ pA; $V_{rms} = 24$ mV) and the corresponding LUMO+1 DFT-LDOS map (right panel). **d**, Laplace-filtered LUMO+1 d$I$/d$V$ map of **1**. The dashed ovals in **c,d** highlight the characteristic

two-lobed LUMO+1 LDOS feature along the edges. **e**, High-resolution STM image of **1** acquired with a CO-functionalized tip ($V = -330$ mV, $I = 100$ pA). Scale bar: 0.5 nm.

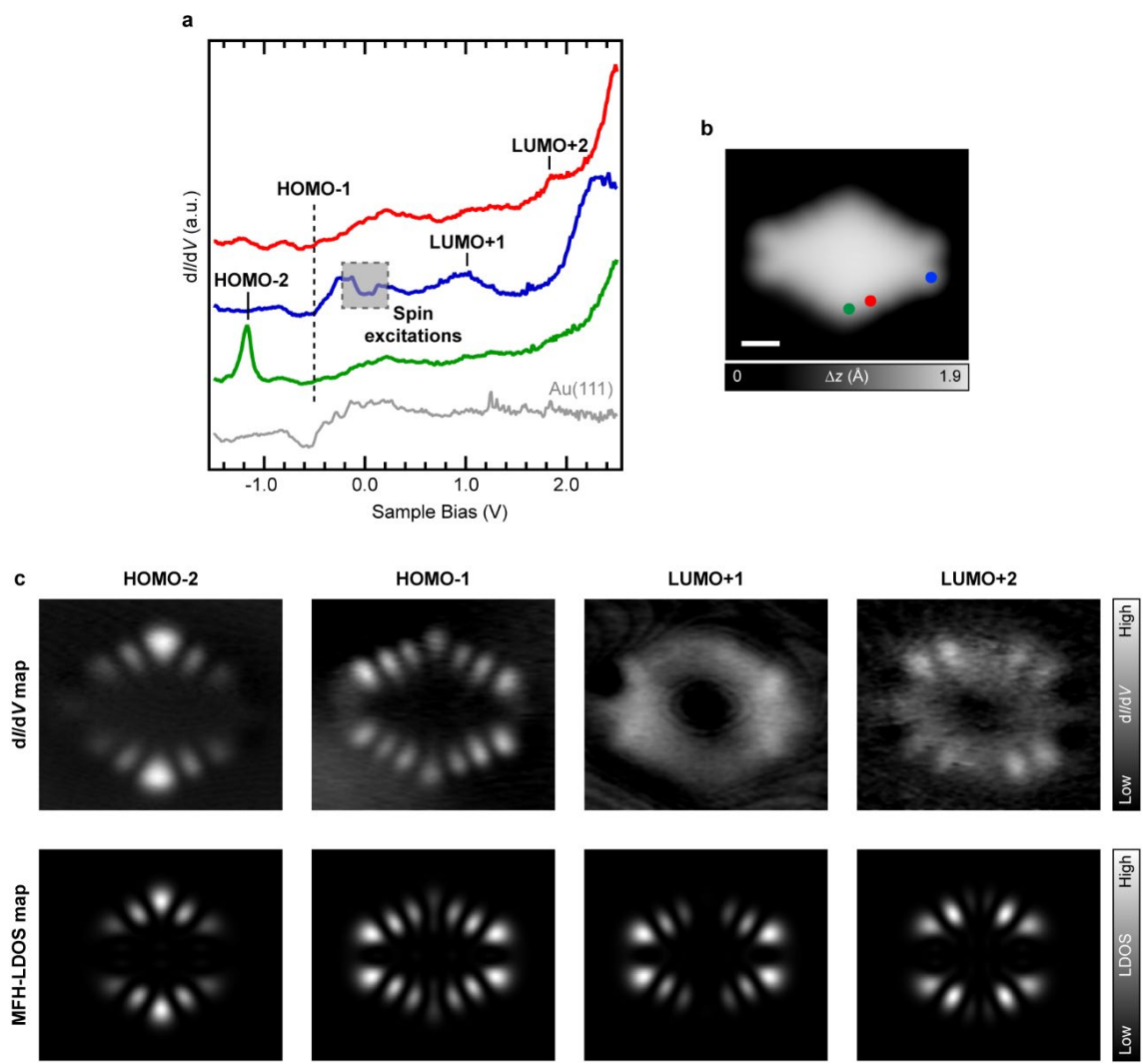

**Supplementary Figure 3 | Detection of non-frontier molecular orbitals of 2 on Au(111). a**, Long-range d$I$/d$V$ spectra acquired on **2**. Acquisition positions are marked with filled circles in **b**. Apart from the spin excitation steps at ±102 mV (gray shaded area), HOMO−2, LUMO+1 and LUMO+2 resonances are detected at −1.17 V, +1.00 V and +1.75 V, respectively. Open feedback parameters: $V = -1.5$ V, $I = 400$ pA; $V_{rms} = 16$ mV. **b**, High-resolution STM image of **2** ($V = -500$ mV, $I = 400$ pA). **c**, Constant-current d$I$/d$V$ maps (upper panels) and MFH-LDOS maps (lower panels) of the HOMO−2 to LUMO+2 resonances of **2** calculated at $U = t_1$. While no clear signature of the HOMO−1 resonance of **2** is found in d$I$/d$V$ spectroscopy, d$I$/d$V$ mapping clearly resolves the HOMO−1 state at −500 mV, close to the onset of the Au(111) surface state. Presumably, the proximity of the HOMO−1 state to the Au(111) surface state obscures the detection of a well-defined HOMO−1 peak in d$I$/d$V$ spectroscopy. Scanning parameters for d$I$/d$V$ maps: $V = -1.17$ V (HOMO−2), −500 mV (HOMO−1), +1.10 V (LUMO+1), +1.75 V (LUMO+2); $I = 400$ pA; $V_{rms} = 22$ mV. Scale bar: 0.5 nm.

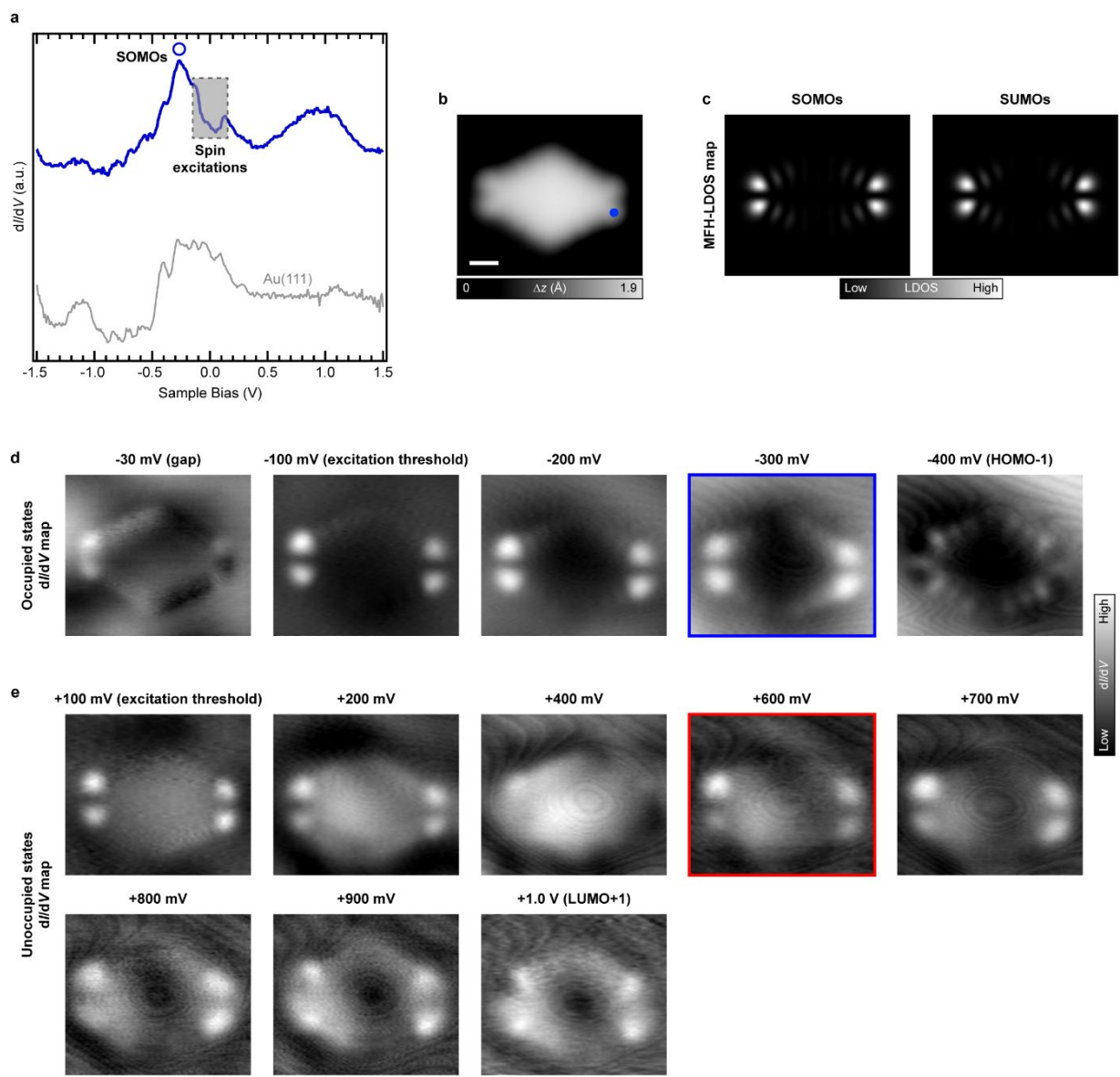

**Supplementary Figure 4 | Detection of frontier molecular orbitals of 2 on Au(111). a**, Long-range d$I$/d$V$ spectrum acquired on **2**. Acquisition position is marked with a filled circle in **b**. The peaked feature at ~ −300 mV is ascribed to the SOMOs resonance. Open feedback parameters: $V = -1.5$ V, $I = 500$ pA; $V_{rms} = 16$ mV. **b**, High-resolution STM image of **2** ($V = -500$ mV, $I = 400$ pA). **c**, MFH-LDOS maps of the SOMOs and SUMOs of **2** calculated at $U = t_1$. **d**, Series of constant-current d$I$/d$V$ maps between −30 mV and −400 mV. LDOS features corresponding to SOMOs emerge at −300 mV (highlighted in blue). **e**, Series of constant-current d$I$/d$V$ maps between +100 mV and +1.0 V. LDOS features corresponding to SUMOs emerge between +600 mV and +900 mV (onset at +600 mV is highlighted in red). Scanning parameters for d$I$/d$V$ maps: $I = 300$ pA (map at −30 mV) and 400 pA (all other maps); $V_{rms} = 12$ mV (map at −30 mV) and 22 mV (all other maps). Scale bar: 0.5 nm.

**Supplementary Note 1. Additional STS measurements on 2 on Au(111).** As shown in Supplementary Figs. 3 and 4, while d$I$/d$V$ spectroscopy on **2** provides clear identification of the majority of non-frontier states, no clear signatures of the frontier states (that is, SOMOs and SUMOs) are found. As shown in Supplementary Fig. 4a, while a peak at ~ −300 mV may be ascribed to SOMOs, SUMOs present no distinguishable features. We therefore acquire a series of d$I$/d$V$ maps to unambiguously identify the frontier orbital resonances. Supplementary Fig. 4d,e present a series of d$I$/d$V$ maps between the spin excitation threshold, and the first non-frontier orbital resonance at each bias polarity (that is, HOMO−1/LUMO+1). Characteristic LDOS features emerge at −300 mV (Supplementary Fig. 4d), and between +600 mV and +900 mV (Supplementary Fig. 4e), which exhibit excellent correspondence to the MFH-LDOS maps of the SOMOs and SUMOs of **2** (Fig. Supplementary Fig. 4c), respectively. Based on d$I$/d$V$ spectroscopy and the onset of the LDOS features in d$I$/d$V$ maps, we estimate the SOMOs and SUMOs resonances to be at −300 mV and +600 mV, respectively, yielding a Coulomb gap of 900 meV. The excellent agreement between the experimental d$I$/d$V$ maps and the gas-phase MFH-LDOS maps evidences the lack of any significant perturbation of the molecular orbitals of **2** when adsorbed on Au(111).

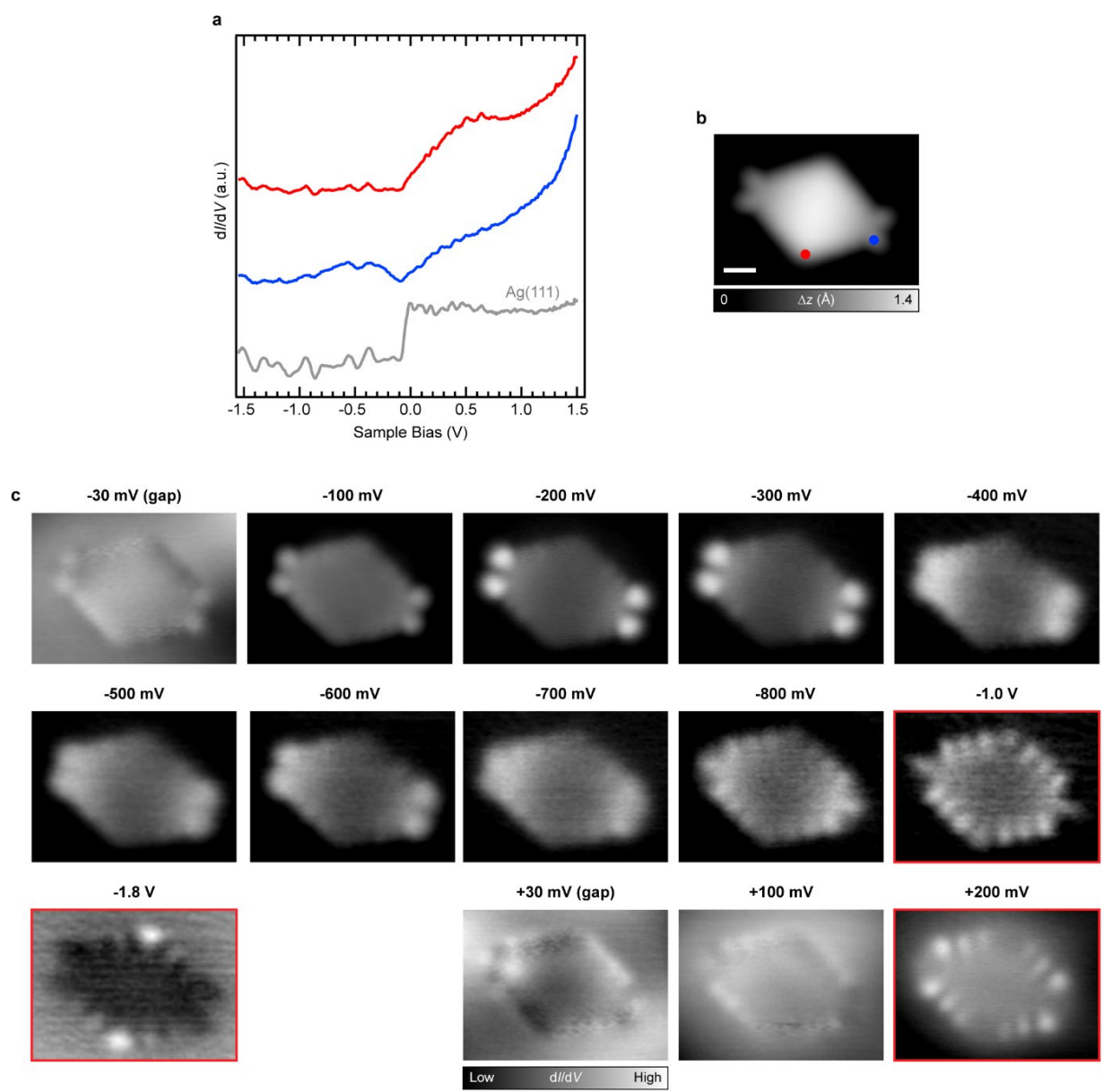

**Supplementary Figure 5 | STS measurements on 2 on Ag(111). a**, Long-range d$I$/d$V$ spectra acquired on **2**. Acquisition positions are marked with filled circles in **b**. For **2** on Ag(111), no clear signatures of molecular orbitals are found with d$I$/d$V$ spectroscopy. Open feedback parameters: $V = -1.5$ V, $I = 400$ pA; $V_{rms} = 16$ mV. **b**, High-resolution STM image of **2** ($V = -300$ mV, $I = 400$ pA). **c**, Series of constant current d$I$/d$V$ maps acquired between −1.8 V and +200 mV. LDOS features corresponding to the HOMO−2 (−1.8 V), HOMO−1 (−1.0 V) and LUMO+1 (+200 mV) of the free-standing molecule are highlighted in red. Scanning parameters for d$I$/d$V$ maps: $I = 250$ pA (maps at ±30 mV) and 400 pA (all other maps); $V_{rms} = 12$ mV (maps at ±30 mV) and 22 mV (all other maps). Scale bar: 0.5 nm.

**Supplementary Note 2. STS measurements on 2 on Ag(111).** We also explored the electronic structure of **2** on Ag(111) with the aim of probing any change in the energetics of magnetic coupling. The on-surface synthesis of **2** on Ag(111) was performed by deposition of **4** on Ag(111), and annealing the surface to 350 °C, which yields both individual molecules and covalently coupled molecular clusters. d$I$/d$V$ spectroscopy on **2** (Supplementary Fig. 5a) provides no clear signatures of molecular orbital resonances. Therefore, we employ sequential d$I$/d$V$ mapping to identify molecular orbital resonances of **2** on Ag(111). As shown in Supplementary Fig. 5c, the electronic structure of **2** on Ag(111) is markedly different from that on Au(111). First, in a bias range of −30 mV to −1.8 V, the following salient features are observed: (1) between −100 mV and −700 mV, LDOS features corresponding to the frontier orbitals of free-standing **2** are observed, (2) between −800 mV and −1.0 V, LDOS features corresponding to the HOMO−1 of free-standing **2** are found, and (3) between −1.6 V to −2.0 V (a representative map at −1.8 V is shown), LDOS features corresponding to the HOMO−2 of free-standing **2** are found. Second, d$I$/d$V$ mapping in the vicinity of the Fermi energy does not yield any signal (representative maps at ±30 mV are shown). Third, the first state to emerge at positive bias (at +200 mV) corresponds to the LUMO+1 of free-standing **2**. These observations suggest charge transfer from the Ag(111) surface to **2**, such that the SUMOs of **2** are fully filled and shift below the Fermi energy, consistent with the ~ 570 meV lower work function of Ag(111) compared to Au(111)[1]. In line with the complete filling of SUMOs, we do not observe any spin excitation steps, consistent with a closed-shell electronic ground state of **2** on Ag(111).

## 2. General methods and materials.

All reactions working with air- or moisture- sensitive compounds were carried out under argon atmosphere using standard Schlenk line techniques. Thin layer chromatography (TLC) was done on silica gel coated aluminum sheets with F254 indicator and column chromatography separation was performed with silica gel (particle size 0.063–0.200 mm). Nuclear magnetic resonance (NMR) spectra were recorded on a Bruker Avance 300 MHz spectrometer. Chemical shifts were reported in ppm relative to the residual of solvents ($CD_2Cl_2$, $^1H$: 5.32 ppm, $^{13}C$: 53.84 ppm; $C_2D_2Cl_4$, $^1H$: 6.00 ppm, $^{13}C$: 73.78 ppm; THF-$d_8$, $^1H$: 3.58 ppm, $^{13}C$: 67.57 ppm). Abbreviations: s = singlet, d = doublet, t = triplet, q = quartet, m = multiplet. Coupling constants ($J$) were presented in Hertz. High-resolution mass spectrometry (HRMS) was performed on a SYNAPT G2-Si high resolution time-of-flight mass spectrometer (Waters Corp., Manchester, UK) by matrix-assisted laser desorption/ionization (MALDI), calibrated against poly(ethylene glycol). Melting points were measured with a Büchi B-545 apparatus. Single crystal diffraction data were collected on a STOE IPDS 2T diffractometer with Cu-K$_\alpha$ radiation for all compounds.

All commercially available chemicals were purchased from TCI, Aldrich, Acros, Merck, and other commercial suppliers and used without further purification unless otherwise noted. Bisanthrone (**5**) and 1,4,8,11-tetramethyl-6,13-pentacenedione (**11**) were prepared following reported procedures[2,3].

## 3. Solution synthesis of precursors 3 and 4.

**Synthesis of 7,14-bis(2,6-dimethylphenyl)ovalene (3).** Precursor **3** was synthesized as shown in Supplementary Scheme 1. Bisanthrone (**5**) was initially treated with 2,6-dimethylphenylmagnesium bromide, followed by quenching with glacial acetic acid to provide isomeric mixture of diol intermediate **6′**. **6′** was then reduced with sodium iodide and sodium hypophosphite monohydrate in glacial acetic acid to give 10,10'-bis(2,6-dimethylphenyl)-9,9'-bianthracene (**6**) in 52% yield over two steps. Next, cyclodehydrogenation of **6** with 2,3-dichloro-5,6-dicyano-1,4-benzoquinone (DDQ) and scandium(III) triflate by refluxing in a mixed solvent of chlorobenzene and triflic acid afforded 7,14-bis(2,6-dimethylphenyl)bisanthene (**7**) in 25% yield. Subsequently, **7** was subjected to two-fold Diels-Alder cycloaddition with nitroethylene *in-situ* generated from 2-nitroethanol and phthalic anhydride, followed by elimination of nitrous acid (HONO) to provide 7,14-bis(2,6-dimethylphenyl)ovalene (**3**) in 68% yield.

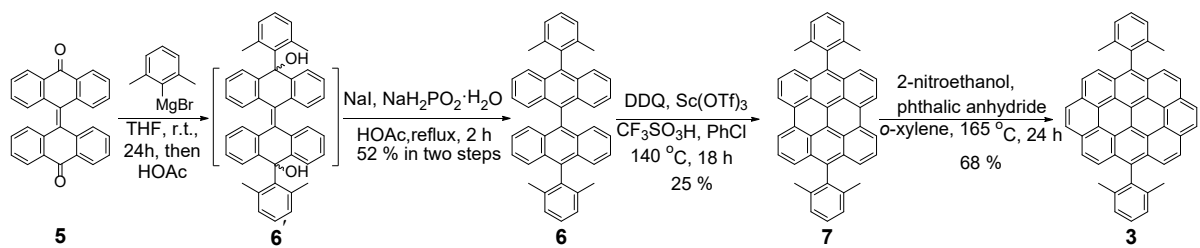

**Supplementary Scheme 1 |** Synthetic route toward 7,14-bis(2,6-dimethylphenyl)ovalene (**3**).

**Synthesis of 10,10'-bis(2,6-dimethylphenyl)-9,9'-bianthracene (6).** To a suspension of **5** (4.61 g, 12.0 mmol) in anhydrous tetrahydrofuran (500 mL) was added a solution of 2,6-dimethylphenylmagnesium bromide in tetrahydrofuran (72 mL, 1.0 M) at room temperature under argon atmosphere. After stirring at room temperature for 24 h, the reaction was quenched by addition of glacial acetic acid (20 mL). The solvent was then evaporated to give a crude product of diol intermediate **6′**, which was used in the next step without further purification. Sodium iodide (13.6 g, 90.7 mmol), sodium hypophosphite monohydrate (14.2 g, 137 mmol), and glacial acetic acid (250 mL) were added to the obtained crude product of **6′**. The mixture was heated under reflux for 4 h. After cooling down to room temperature, water (300 mL) was added and the reaction mixture was extracted with dichloromethane (250 mL) for three times. The organic phases were combined, washed with water and saturated aqueous solution of $NaHCO_3$, and dried over anhydrous $Na_2SO_4$. After the solvent was evaporated, the residue was purified by column chromatography (*n*-hexane:dichloromethane = 4 : 1) to give the title compound (3.54 g, 52% yield) as yellow solid. M.p.: 338.3 – 340.2 °C; $^1$H NMR (300 MHz, $CD_2Cl_2$, 298 K, ppm) $\delta$ 7.62 (d, *J* = 8.7 Hz, 4H), 7.48 – 7.29 (m, 10H), 7.19 (d, *J* = 4.8 Hz, 8H), 1.96 (s, 12H); $^{13}$C NMR (75 MHz, $CD_2Cl_2$, 298 K, ppm) $\delta$ 138.4, 138.3, 137.1, 133.6, 132.2, 130.5, 128.4, 128.1, 127.7, 126.8, 126.3, 126.2, 118.8, 20.5; MS (MALDI-TOF, positive) *m/z*: Calcd for $C_{44}H_{34}$: 562.27; Found: 562.27 [M]$^+$.

**Synthesis of 7,14-bis(2,6-dimethylphenyl)bisanthene (7).** DDQ (1.28 g, 5.64 mmol), Sc(OTf)$_3$ (3.85 g, 7.82 mmol), and triflic acid (5.25 mL) were added to a solution of 10,10'-bis(2,6-dimethylphenyl)-9,9'-bianthracene (**6**) (645 mg, 1.15 mmol) in chlorobenzene (168 mL) and the mixture was heated under reflux for 19 h. After cooling down to room temperature, a solution of hydrazine monohydrate in tetrahydrofuran (12 mL, 1.0 M) was added and the reaction mixture was stirred for 10 minutes. The solution was washed with water and brine, dried over anhydrous $Na_2SO_4$, and evaporated. The residue was purified by column chromatography (*n*-hexane:dichloromethane = 6 : 1) to give the title compound (156 mg, 25% yield) as blue solid. M.p.: > 400 °C, decomposed; $^1$H NMR (300 MHz, THF-$d_8$, 298 K, ppm) $\delta$ 8.39 (d, *J* = 7.5 Hz, 4H), 7.41 – 7.26 (m, 10H), 7.16 (d, *J* = 8.6 Hz, 4H), 1.91 (s, 12H); $^{13}$C NMR (75 MHz, THF-$d_8$, 298 K, ppm) $\delta$ 138.5, 135.8, 133.8, 132.6, 128.9, 128.8, 128.3, 127.5, 126.4, 121.7, 20.9; MS (MALDI-TOF, positive) *m/z*: Calcd for $C_{44}H_{30}$: 558.23; Found: 558.25 [M]$^+$.

**Synthesis of 7,14-bis(2,6-dimethylphenyl)ovalene (3).** To an oven-dried Schlenk tube was added 7,14-bis(2,6-dimethylphenyl)bisanthene (**7**) (56 mg, 0.10 mmol), followed by 2-nitroethanol (911 mg, 10.0 mmol) and dry *o*-xylene (10 mL) via syringe. After degassing by three freeze-pump-thaw cycles with argon, phthalic anhydride (1.48 g, 10.0 mmol) was added and the mixture was heated at 165 °C for 24 h. After cooling down the mixture to room temperature, the solvent was evaporated. The obtained residue was purified by column chromatography (*n*-hexane:dichloromethane = 5 : 1) to give title compound (41 mg, 68% yield) as orange solid, which was recrystallized for 4 times from dichloromethane and methanol before the use for the surface synthesis. M.p.: > 400 °C, decomposed; $^1$H NMR (300 MHz, $C_2D_2Cl_4$, 298 K, ppm) δ 9.34 (s, 4H), 9.04 (d, *J* = 8.9 Hz, 4H), 8.74 (d, *J* = 6.1 Hz, 4H), 7.68 – 7.54 (m, 6H), 1.94 (s, 12H); $^{13}$C NMR (75 MHz, $C_2D_2Cl_4$, 298 K, ppm) δ 138.4, 138.3, 134.1, 128.5, 128.1, 128.0, 127.8, 127.0, 125.5, 125.3, 121.9, 120.4, 20.9; HRMS (MALDI-TOF, positive) *m/z*: Calcd for $C_{48}H_{30}$: 606.2342; Found: 606.2343 [M]$^+$.

**Synthesis of 6,13-bis{10-(2,6-dimethylphenyl)anthracen-9-yl}-1,4,8,11-tetramethylpentacene (4).** Precursor **4** was synthesized as shown in Supplementary Scheme 2. 9-(2,6-Dimethylphenyl)anthracene (**9**) was initially prepared through Suzuki coupling reaction of 9-bromoanthracene (**8**) with (2,6-dimethylphenyl)boronic acid in 73% yield. Then, bromination of **9** with bromine provided 9-bromo-10-(2,6-dimethylphenyl)anthracene (**10**) in 96% yield. Subsequently, **11** was reacted with {10-(2,6-dimethylphenyl)anthracen-9-yl}lithium generated by lithiation of **10**, followed by dehydroxylation with sodium iodide/sodium hypophosphite monohydrate to give precursor **4** in 29% yield over two steps.

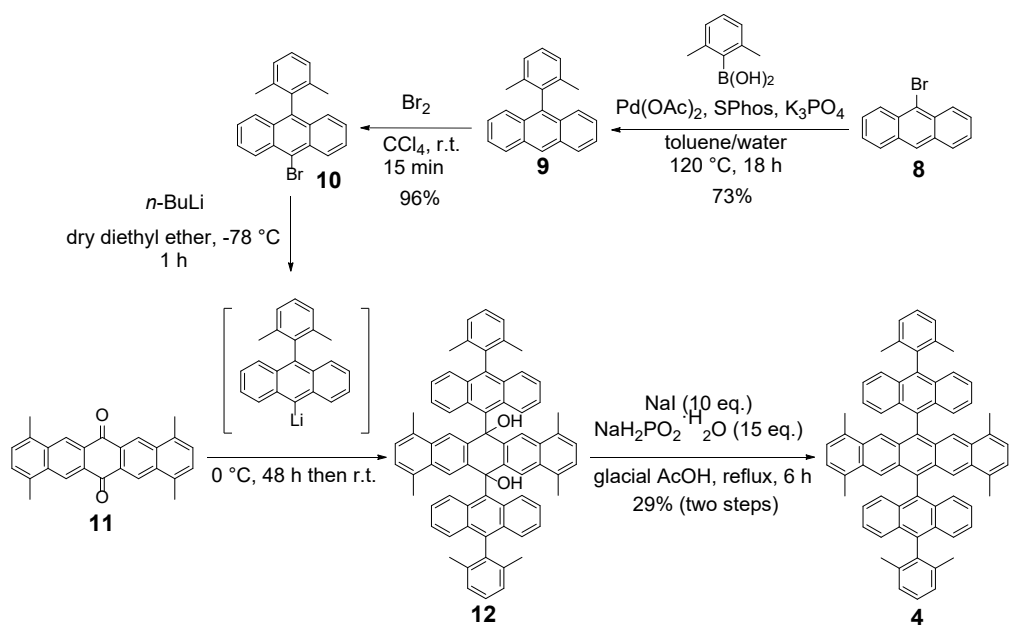

**Supplementary Scheme 2** | Synthetic route toward 6,13-bis{10-(2,6-dimethylphenyl)anthracen-9-yl}-1,4,8,11-tetramethylpentacene (**4**).

**Synthesis of 9-(2,6-dimethylphenyl)anthracene (9).** 9-Bromoanthracene (3.00 g, 11.7 mmol), 2,6-dimethylphenylboronic acid (3.50 g, 23.3 mmol), Pd(OAc)$_2$ (78.9 mg, 0.35 mmol), 2-dicyclohexylphosphino-2',6'-dimethoxybiphenyl (SPhos) (289 mg, 0.700 mmol), and $K_3PO_4$ (6.19 g, 29.2 mmol) were placed in a dried flask and toluene/water (100 mL/10 mL) was added. The resulting mixture was subjected to freeze-pump-thaw cycles (3 times) and heated at 120 °C under stirring for 18 h. After cooling the reaction mixture to room temperature, water (60 mL) was added, and the aqueous layer was extracted with dichloromethane for three times. Combined organic layers were washed with brine, dried over anhydrous MgSO$_4$, filtered, and concentrated in vacuo. The resulting dark reddish-brown semi-solid was dissolved in a minimal amount of dichloromethane and added to MeOH (50 mL). The desired product **9** precipitated as a white solid, which was filtered, washed with ice-cold MeOH, and then dried under vacuum to obtain 2.40 g of **9** (73% yield). M.p.: 146.5–147.8 °C. $^1$H NMR (300 MHz, C$_2$D$_2$Cl$_4$, 298 K, ppm) $\delta$ 8.52 (s, 1H), 8.09 (d, $J$ = 8.4 Hz, 2H), 7.55 – 7.43 (m, 4H), 7.37 (q, $J$ = 8.4 Hz, 3H), 7.29 (d, $J$ = 7.5 Hz, 2H), 6.00 (s, 1H), 1.76 (s, 6H); $^{13}$C NMR (75 MHz, C$_2$D$_2$Cl$_4$, 298 K, ppm) $\delta$ 137.57, 137.30, 135.39, 131.35, 129.30, 128.49, 127.55, 127.35, 126.08, 125.73, 125.60, 125.16, 20.02. HRMS (MALDI-TOF, positive) $m/z$: Calcd for C$_{22}$H$_{18}$: 282.1409; Found: 282.1403 [M]$^+$.

**Synthesis of 9-bromo-10-(2,6-dimethylphenyl)anthracene (10).** To an oven-dried 100-mL flask was added **9** (1.40 g, 4.96 mmol) and CCl$_4$ (22 mL). Bromine (0.832 g, 0.268 mL, 5.21 mmol) was then added dropwise over 5 minutes. The reaction mixture was stirred at room temperature for 15 min under the exclusion of light and then quenched with saturated aqueous solution of Na$_2$SO$_3$ (50 mL). The mixture was extracted with dichloromethane for three times and washed with water and brine. The organic layer was then dried over anhydrous MgSO$_4$, filtered, and concentrated in vacuo. Purification by recrystallization from chloroform/methanol afforded the title compound **10** (1.71 g, 96% yield) as dark green solid. M.p.: 167.3–168.5 °C. $^1$H NMR (300 MHz, C$_2$D$_2$Cl$_4$, 298 K, ppm) $\delta$ 8.62 (d, $J$ = 8.9 Hz, 2H), 7.63 (t, $J$ = 7.0 Hz, 2H), 7.49 (d, $J$ = 8.7 Hz, 2H), 7.45 – 7.36 (m, 3H), 7.29 (d, $J$ = 7.5 Hz, 2H), 1.74 (s, 6H); $^{13}$C NMR (75 MHz, C$_2$D$_2$Cl$_4$, 298 K, ppm) $\delta$ 137.42, 136.84, 136.30, 130.24, 130.12, 127.88, 127.85, 127.48, 127.13, 126.16, 125.97, 20.03. HRMS (MALDI-TOF, positive) $m/z$: Calcd for C$_{22}$H$_{17}$Br: 360.0514; Found: 360.0515 [M]$^+$.

**Synthesis of 6,13-bis{10-(2,6-dimethylphenyl)anthracen-9-yl}-1,4,8,11-tetramethylpentacene (4).** To a suspension of **10** (327 mg, 0.905 mmol) in dry diethyl ether (40 mL) was added a solution of *n*-butyllithium (*n*-BuLi) (0.57 mL, 0.91 mmol, 1.6 M in hexanes) dropwise at –78 °C. The reaction mixture was stirred for 1 h under argon to obtain {10-(2,6-dimethylphenyl)anthracen-9-yl}lithium. To a suspension of 7,10-dimethyltetracene-5,12-dione (**11**) (150 mg, 0.412 mmol) in dry diethyl ether (15 mL) was added at 0 °C the separately prepared lithium reagent via a double-tipped needle under argon. The reaction mixture was stirred at 0 °C for 48 h, and then allowed to gradually warm up to room temperature. Then, the reaction was quenched by adding 5 mL of glacial acetic acid. The precipitates were collected by filtration and washed with diethyl ether. To the resulting crude material placed in a 100-mL round-bottom flask was added glacial acetic acid (45 mL), NaI (617 mg, 4.12 mmol) and NaH$_2$PO$_2$·H$_2$O (654 mg, 6.17 mmol). The reaction mixture was refluxed for 6 h under the exclusion of light. After cooling down to the room temperature, the precipitations were collected by filtration, and then washed with water and methanol to afford monomer **4** as dark blue solid (106 mg, 29% yield). For the on-surface experiments, further purification was performed by recrystallization through slow diffusion of degassed methanol into a degassed solution of **4** in 1,2-dichlorobenzene under argon. M.p.: >300 °C. $^1$H NMR (300 MHz, C$_2$D$_2$Cl$_4$, 298 K, ppm) $\delta$ 8.13 (s, 4H), 7.72 (d, $J$ = 8.7 Hz, 4H), 7.43 (m, 14H), 7.24 (t, $J$ = 7.6 Hz, 4H), 6.88 (s, 4H), 2.06 (s, 12H), 2.02 (s, 12H); $^{13}$C NMR (75 MHz, C$_2$D$_2$Cl$_4$, 298 K, ppm) $\delta$ 137.64, 132.33, 131.66, 131.05, 130.08, 129.45, 129.34, 129.02, 128.19, 127.57, 127.34, 126.42, 125.28, 125.23, 19.85, 18.53. HRMS (MALDI-TOF, positive) $m/z$: Calcd for C$_{48}$H$_{30}$Br$_2$: 894.4226; Found: 894.4210 [M]$^+$.

# 4. Characterization data of chemical compounds.

## 4.1. Single-crystal X-ray diffraction analysis.

Single crystals of 7,14-bis(2,6-dimethylphenyl)ovalene (**3**) suitable for X-ray diffraction analysis were obtained by slow diffusion of acetonitrile to its solution in 1,1,2,2-tetrachloroethane as light brown needles. The structure was deposited at Cambridge Crystallographic Data Centre (CCDC number: 1978171).

*Crystal data*

| | |
|---|---|
| formula | $C_{48}H_{30}$ |
| molecular weight | 606.75 gmol$^{-1}$ |
| absorption | $\mu = 0.57$ mm$^{-1}$ |
| crystal size | 0.03 x 0.05 x 0.44 mm$^3$ light brown needle |
| space group | P 2$_1$/c (monoclinic) |
| lattice parameters | a = 7.5004(9) Å |
| (calculate from | b = 14.0537(16) Å, ß = 102.809(9)° |
| 6010 reflections with | c = 14.7690(17) Å |
| 3.07° < θ < 64.18°) | V = 1518.0(3) Å$^3$, z = 2, F(000) = 636 |
| temperature | -80 °C |
| density | $d_{xray} = 1.327$ gcm$^{-3}$ |

*Data collection*

| | |
|---|---|
| diffractometer | STOE IPDS 2T |
| radiation | Cu-K$_\alpha$ IµS mirror system |
| Scan – type | ω scans |
| Scan – width | 1° |
| scan range | 2° ≤ θ < 68.1° |
| | -9 ≤ h ≤ 9   -16 ≤ k ≤ 16   -17 ≤ l ≤ 17 |
| number of reflections: | |
| measured | 8972 |
| unique | 2698 ($R_{int}$ = 0.0498) |
| observed | 1143 (|F|/σ(F) > 4.0) |

*Data correction, structure solution and refinement*

| | |
|---|---|
| corrections | Lorentz and polarisation correction. |
| Structure solution | Program: SIR-2004 (Direct methods) |
| refinement | Program: SHELXL-2018 (full matrix). 220 refined parameters, weighting scheme: |
| | w=1/[σ$^2$(F$_o^2$) + (0.1662*P)$^2$+3.29*P] |
| | with (Max(F$_o^2$,0)+2*F$_c^2$)/3. H-atoms at calculated positions and refined with isotropic displacement parameters, non H- atoms refined anisotropically. |
| R-values | wR2 = 0.4011 (R1 = 0.1204 for observed reflections, 0.2246 for all reflections) |
| goodness of fit | S = 1.041 |
| maximum deviation of parameters | 0.001 * e.s.d |
| maximum peak height in diff. Fourier synthesis | 0.4, -0.25 eÅ$^{-3}$ |
| remark | molecule has C$_i$ symmetry |

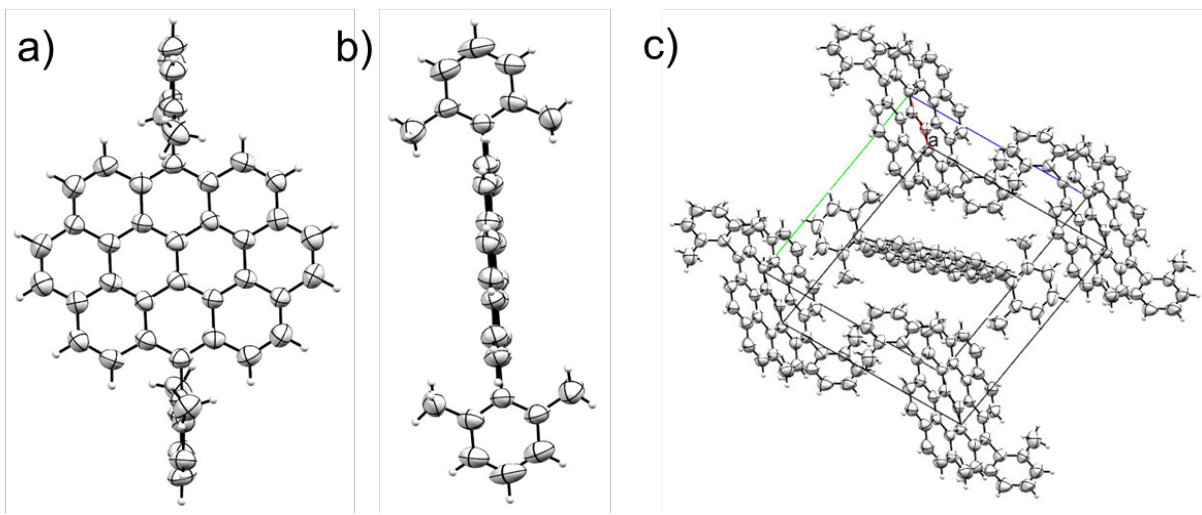

**Supplementary Figure 6 |** X-ray single-crystal structure of **3** with thermal ellipsoids set at 50% probability level. **a**, front view; **b**, side view and **c**, three-dimensional packing mode.

The single crystal of **4** suitable for X-ray analysis was obtained by slow diffusion of degassed methanol into a degassed solution of **4** in 1,2-dichlorobenzene under argon and exclusion of light. The structure was deposited at Cambridge Crystallographic Data Centre (CCDC number: 1978172).

*Crystal data*

| | |
|---|---|
| formula | $C_{70}H_{54}$, $C_6H_4Cl_2$ |
| molecular weight | 1042.12 gmol$^{-1}$ |
| absorption | µ = 1.391 mm$^{-1}$ |
| crystal size | 0.03 x 0.03 x 0.08 mm$^3$ brown needle |
| space group | P $\bar{1}$ (triclinic) |
| lattice parameters | a = 10.4168(11)Å   α = 97.853(8)° |
| (calculate from | b = 11.0736(11)Å   β = 108.311(8)° |
| 9347 reflections with | c = 12.8739(14)Å   γ = 92.478(8)° |
| 3.6° < θ < 66.3°) | V = 1390.7(2)Å$^3$   z = 1   F(000) = 548.0 |
| temperature | 120K |
| density | $d_{xray}$ = 1.244 gcm$^{-3}$ |

*Data collection*

| | |
|---|---|
| diffractometer | STOE IPDS 2T |
| radiation | Cu-K$_α$ IµS mirror system |
| Scan – type | ω scans |
| Scan – width | 1° |
| scan range | 2° ≤ θ < 67.9° |
| | -12 ≤ h ≤ 12   -13 ≤ k ≤ 13   -14 ≤ l ≤ 14 |
| number of reflections: | |
| measured | 19029 |
| unique | 4797 ($R_{int}$ = 0.2244) |
| observed | 2063 (|F|/σ(F) > 4.0) |

*Data correction, structure solution and refinement*

| | |
|---|---|
| corrections | Lorentz and polarisation correction. |
| Structure solution | Program: SIR-2004 (Direct methods) |
| refinement | Program: SHELXL-2018 (full matrix). 360 refined parameters, weighting scheme: $w=1/[\sigma^2(F_o^2) + (0.0959*P)^2+4.21*P]$ with $(Max(F_o^2,0)+2*F_c^2)/3$. H-atoms at calculated positions and refined with isotropic displacement parameters, non H- atoms refined anisotropically. Solvence molecule isotropic refined. |
| R-values | wR2 = 0.3200 (R1 = 0.1022 for observed reflections, 0.2343 for all reflections) |
| goodness of fit | S = 1.080 |
| maximum deviation of parameters | 0.001 * e.s.d |
| maximum peak height in diff. Fourier synthesis | 0.4, -0.4 eÅ$^{-3}$ |
| remark | molecule has $C_i$ symmetry, solvent molecule (1,2-dichlorobenzene is disordered). |

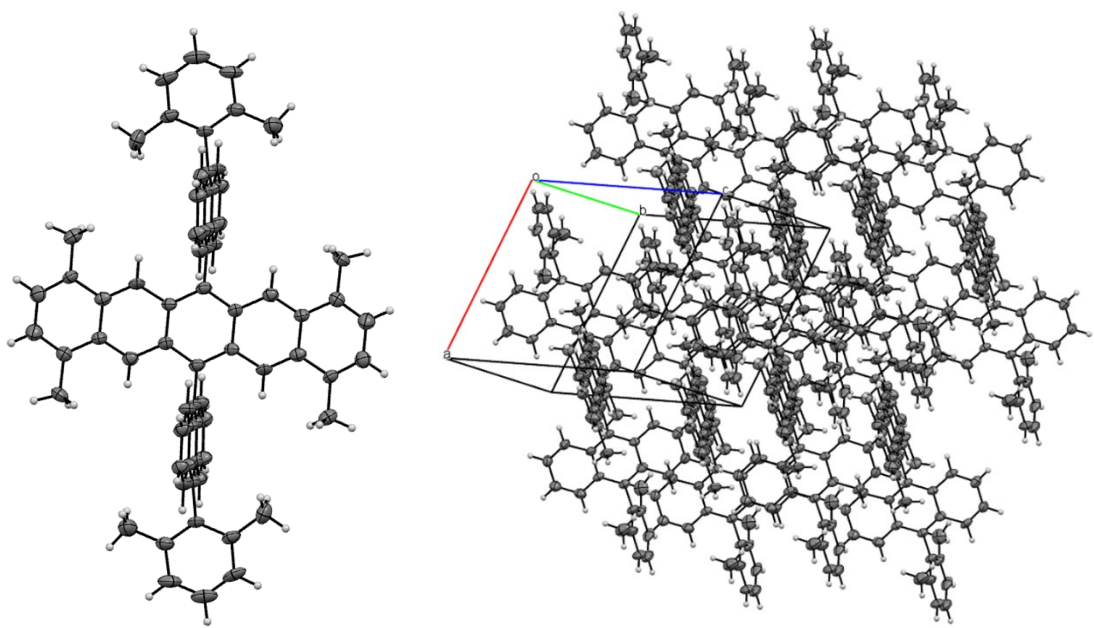

**Supplementary Figure 7** | X-ray single-crystal structure of **4** (ORTEP drawings with thermal ellipsoids set at 50% probability, solvent molecules were removed for clarity).

## 4.2. NMR and mass spectroscopy.

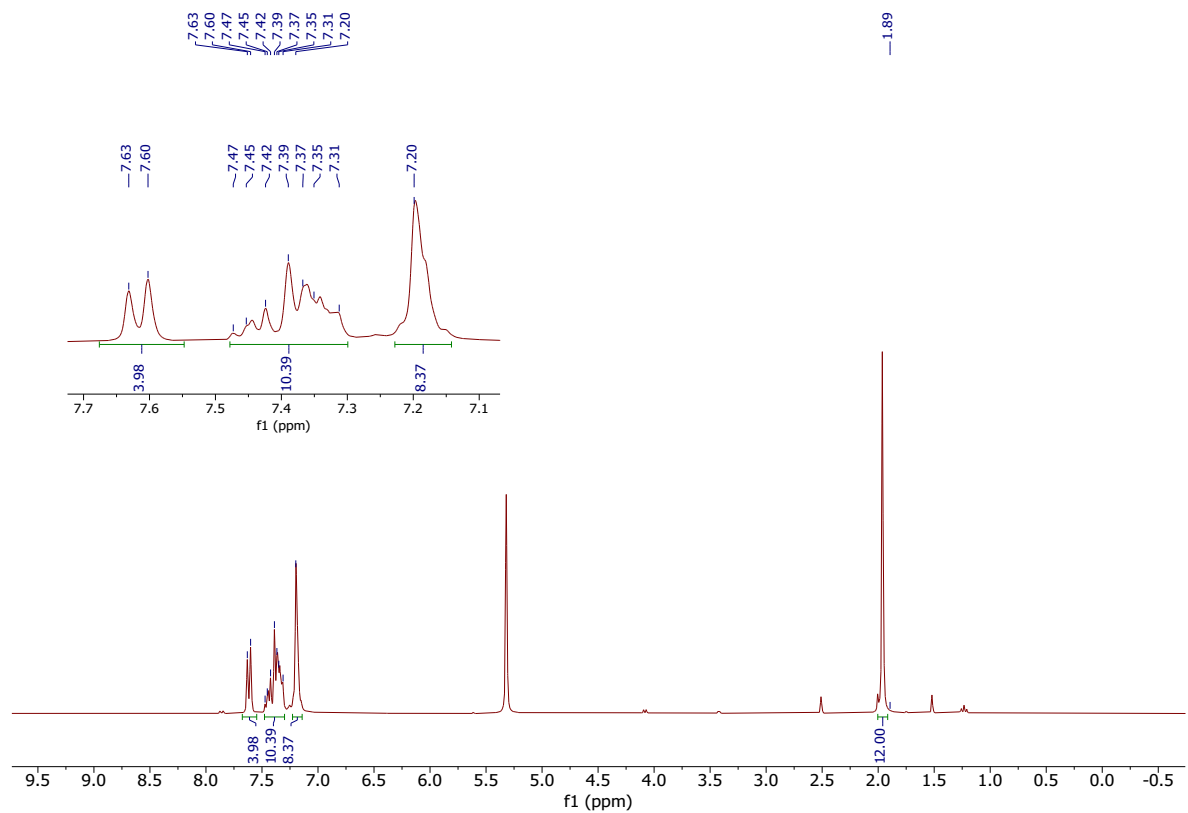

**Supplementary Figure 8** | ¹H NMR spectrum of compound **6** (300 MHz, CD$_2$Cl$_2$, 298 K).

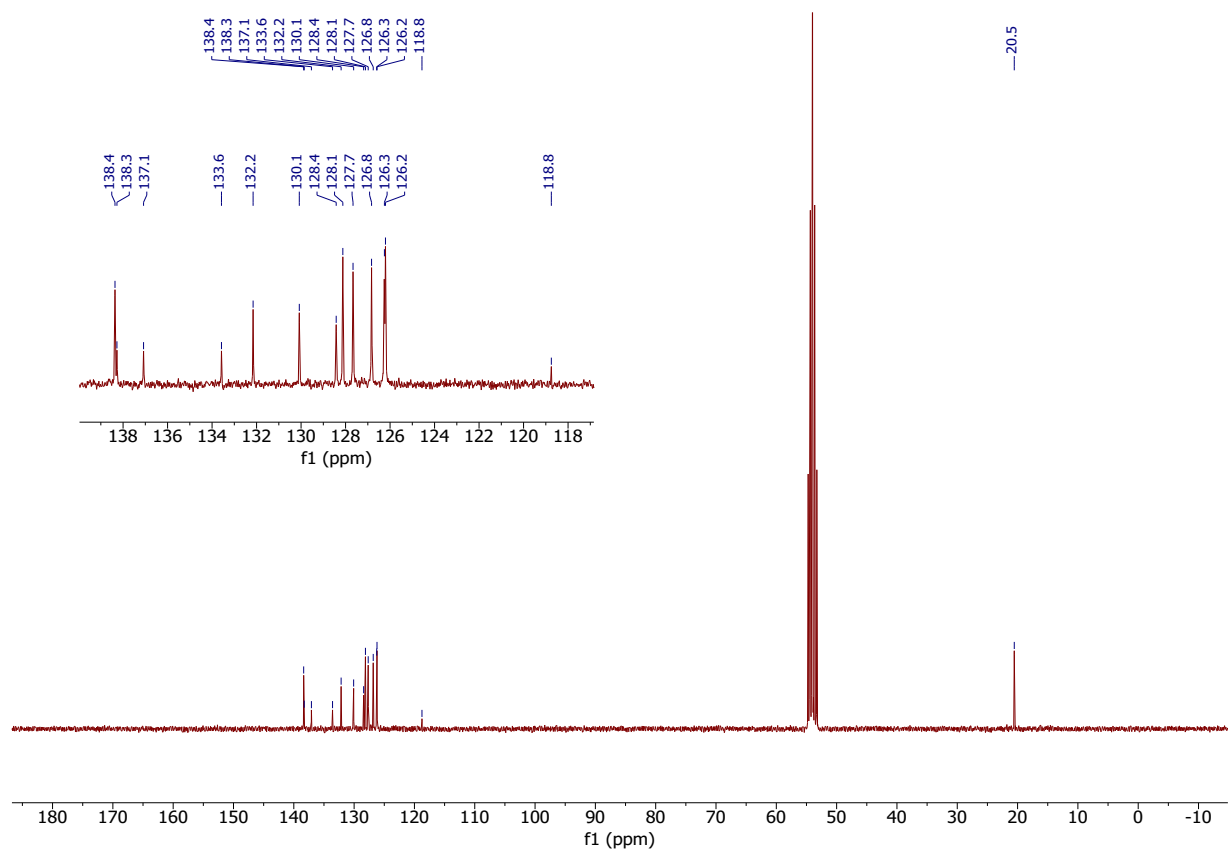

**Supplementary Figure 9** | $^{13}$C NMR spectrum of compound **6** (75 MHz, CD$_2$Cl$_2$, 298 K).

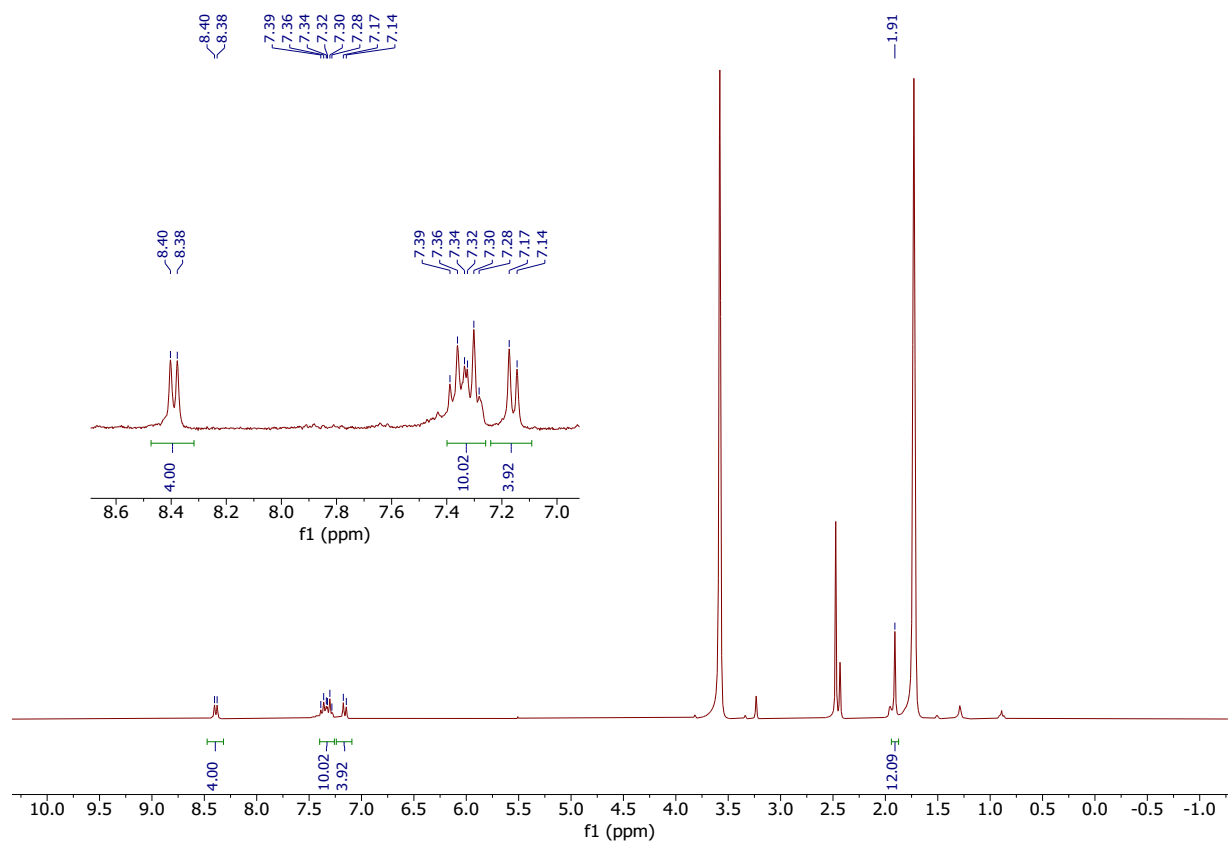

**Supplementary Figure 10** | $^1$H NMR spectrum of compound **7** (300 MHz, THF-$d_8$, 298 K).

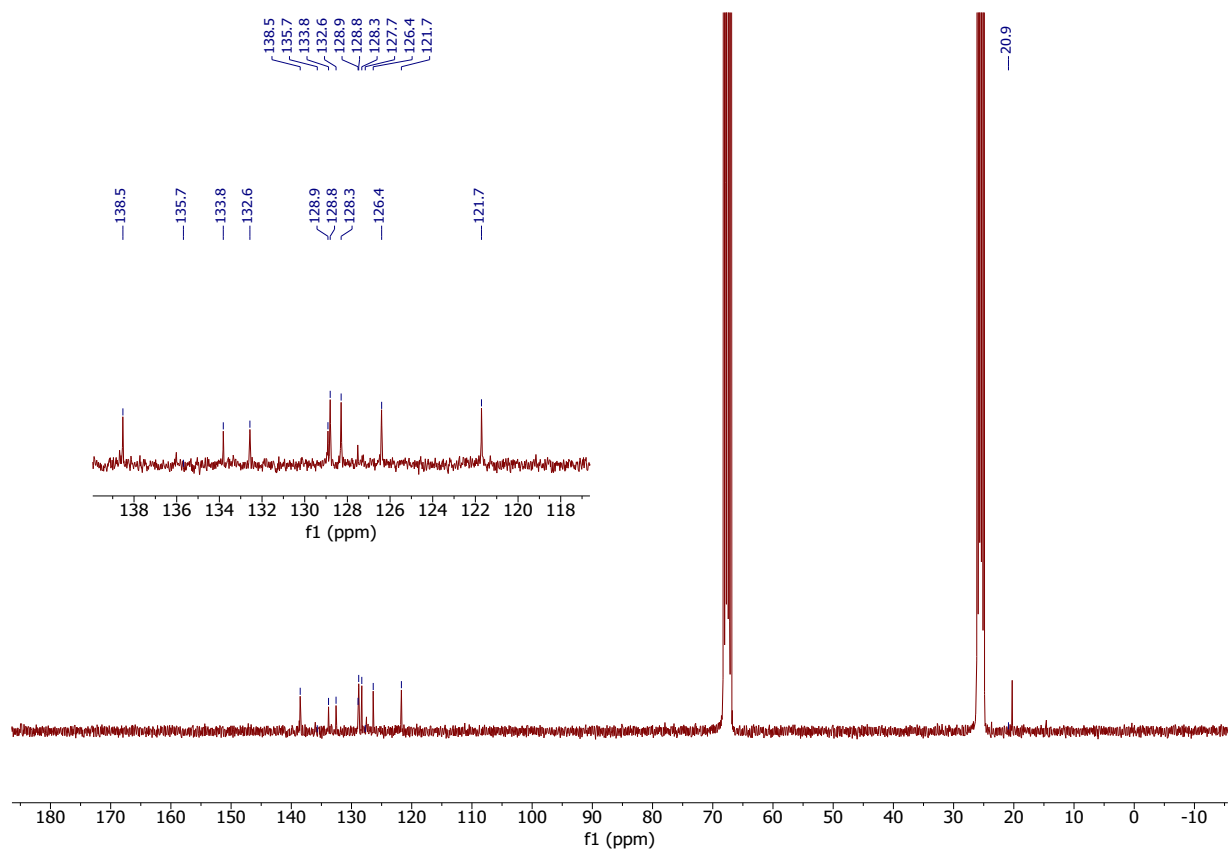

**Supplementary Figure 11** | $^{13}$C NMR spectrum of compound **7** (75 MHz, THF-$d_8$, 298 K).

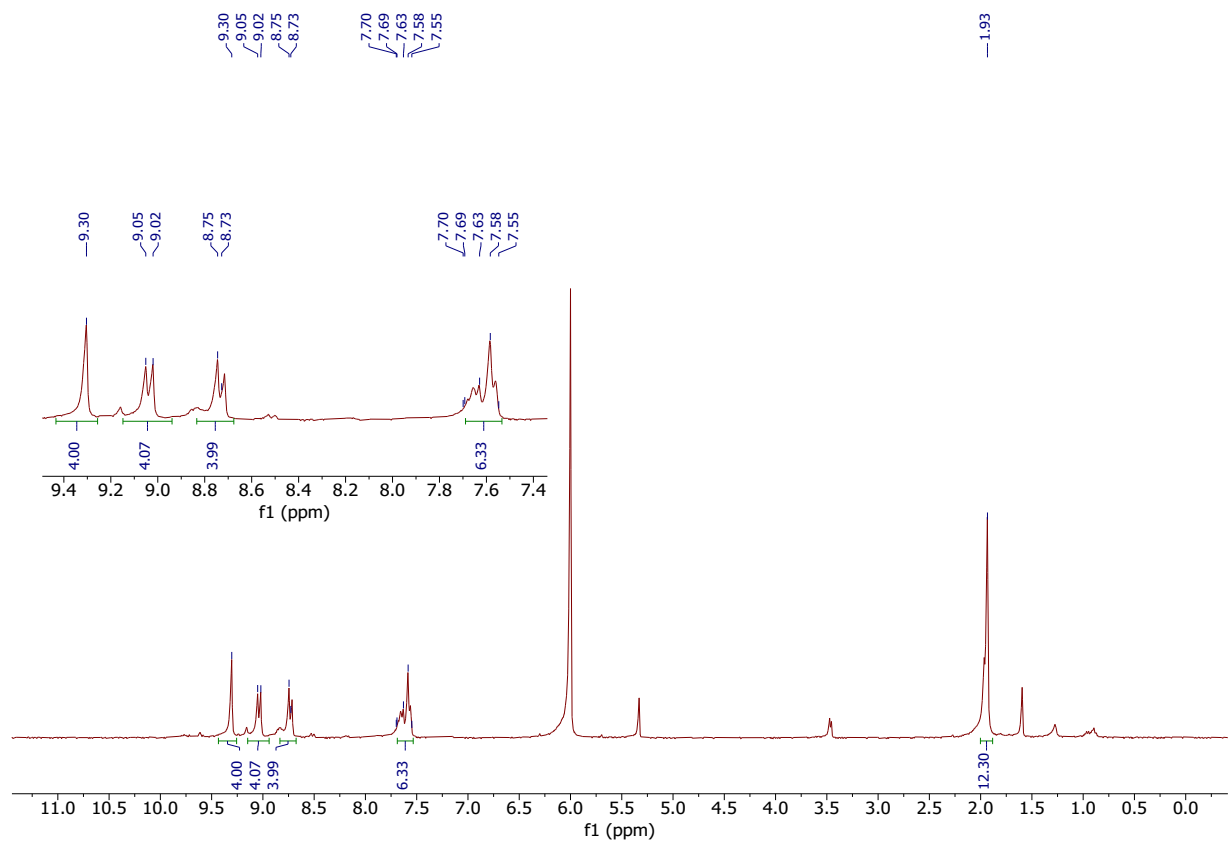

**Supplementary Figure 12** | $^1$H NMR spectrum of precurosr **3** (300 MHz, $C_2D_2Cl_4$, 298 K).

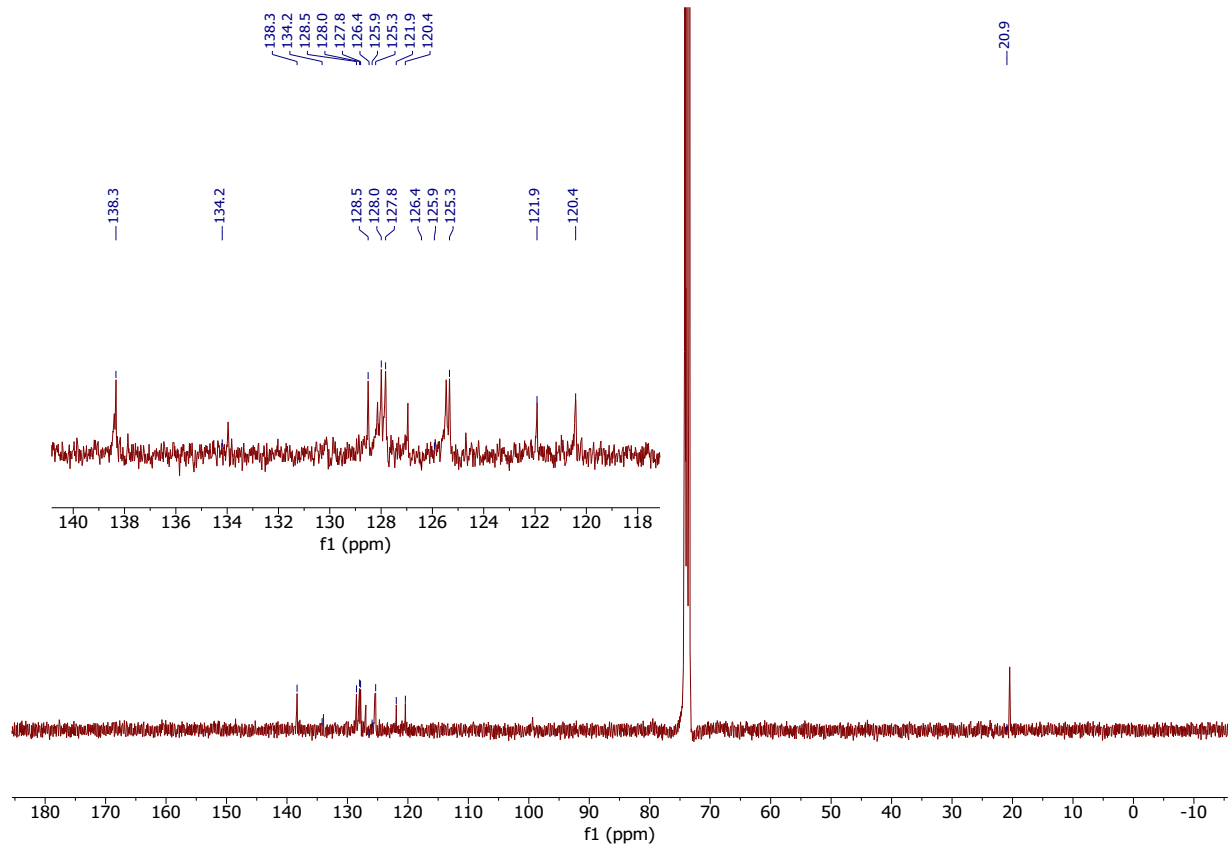

**Supplementary Figure 13** | $^{13}$C NMR spectrum of precursor **3** (75 MHz, $C_2D_2Cl_4$, 298 K).

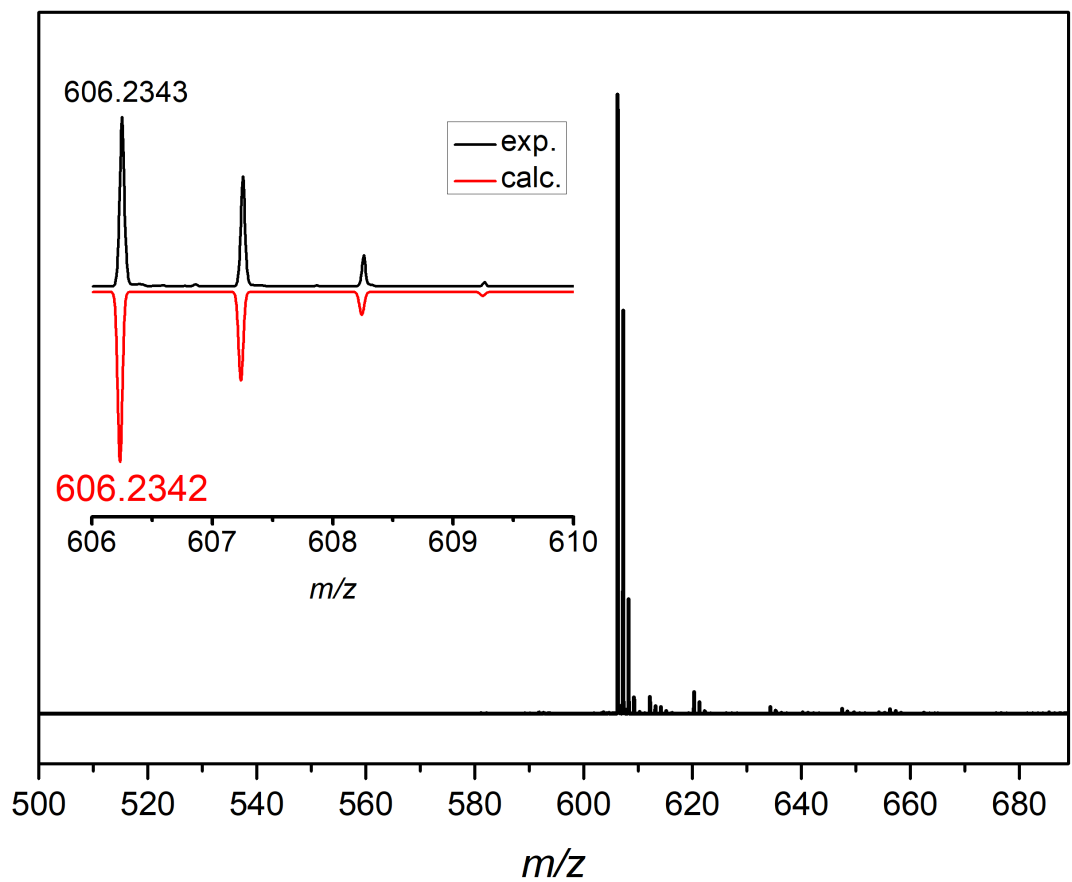

**Supplementary Figure 14 |** MALDI-TOF MS spectra of precursor **3**. Inset shows experimental isotopic distribution pattern in comparison with simulation.

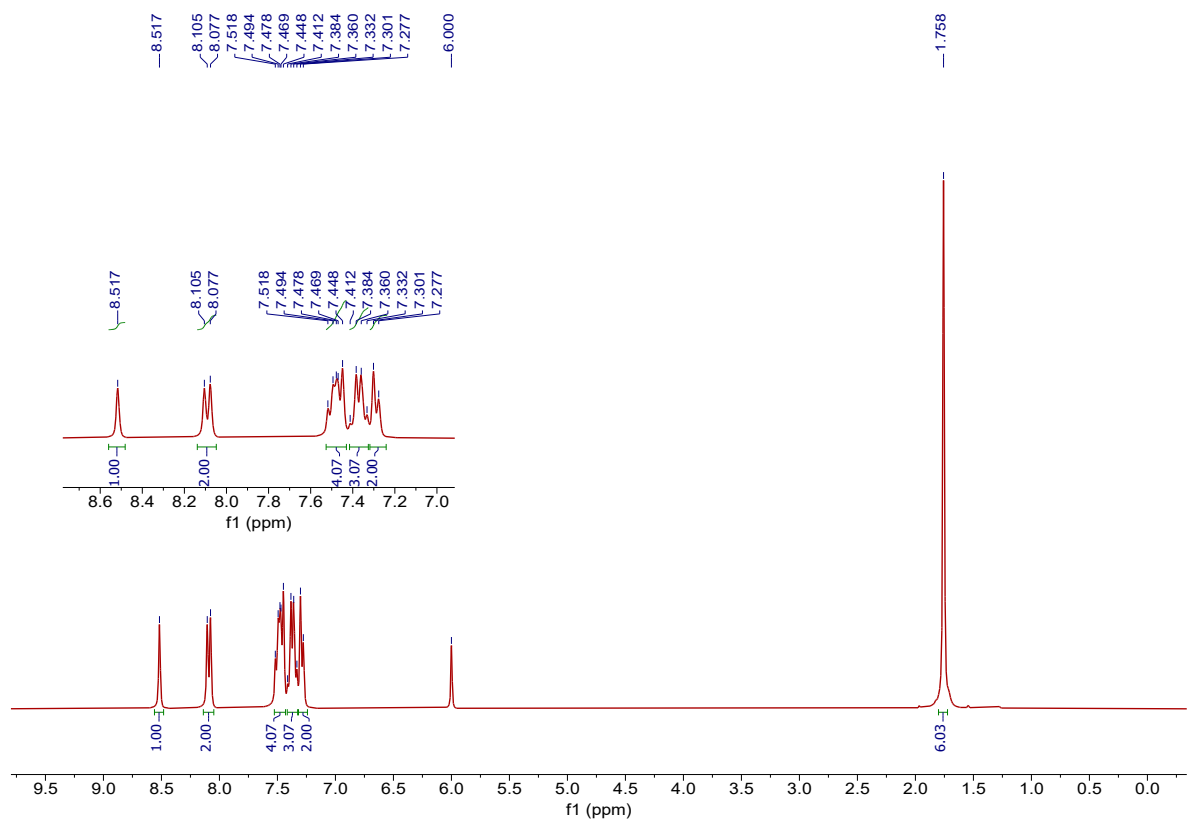

**Supplementary Figure 15 |** $^1$H NMR spectrum of compound **9** (300 MHz, C$_2$D$_2$Cl$_4$, 298 K).

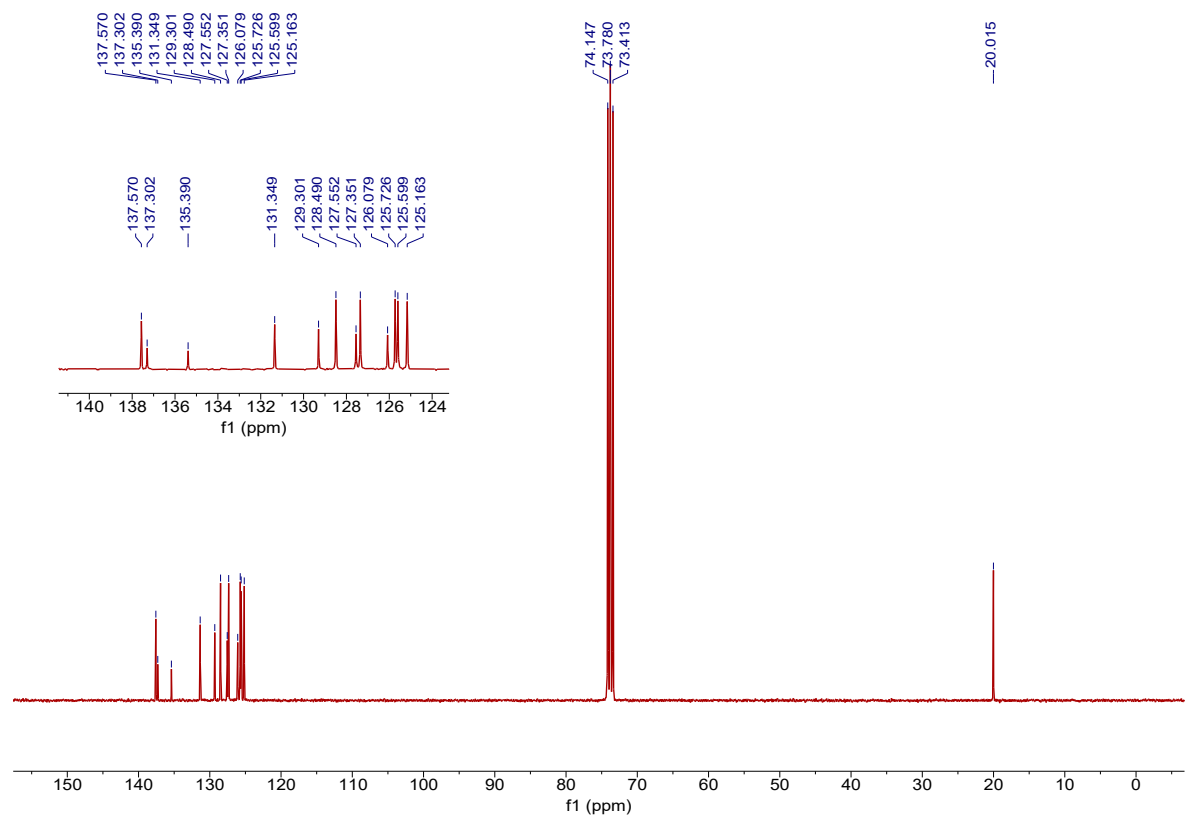

**Supplementary Figure 16 |** $^{13}$C NMR spectrum of compound **9** (75 MHz, C$_2$D$_2$Cl$_4$, 298 K).

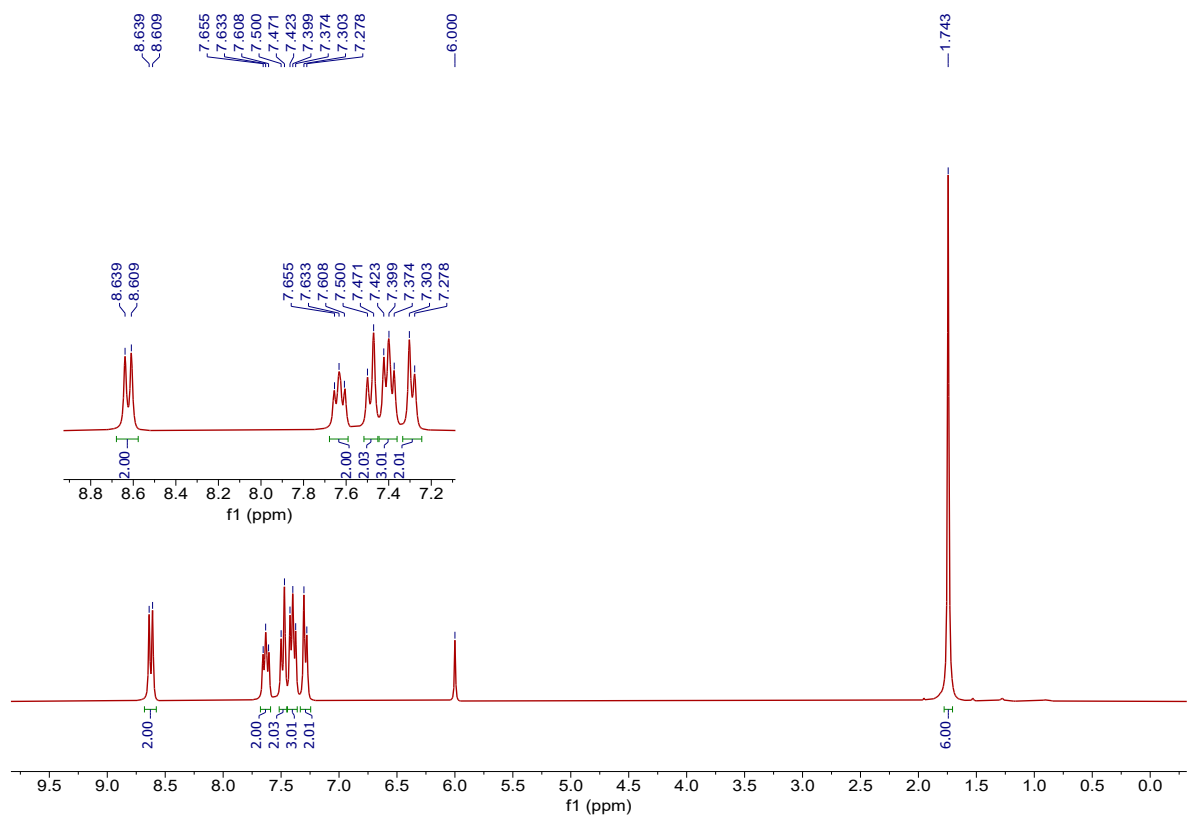

**Supplementary Figure 17** | $^1$H NMR spectrum of compound **10** (300 MHz, C$_2$D$_2$Cl$_4$, 298 K).

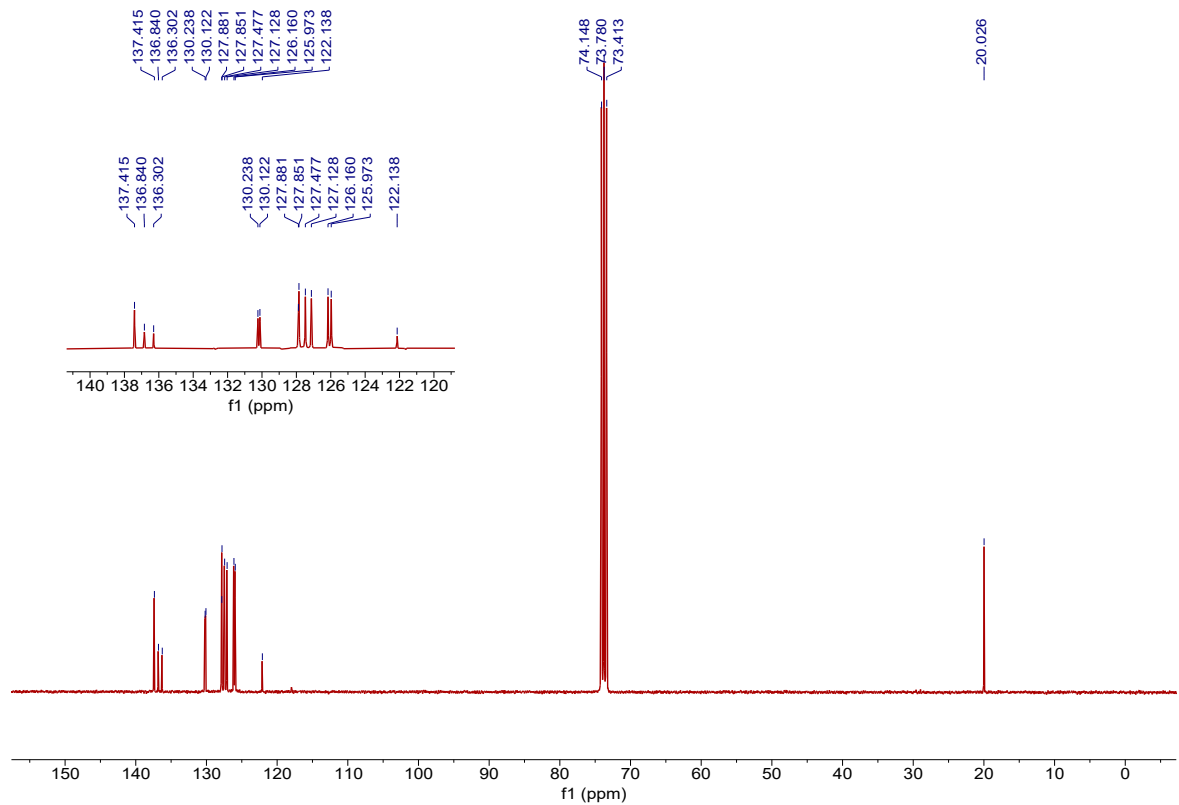

**Supplementary Figure 18** | $^{13}$C NMR spectrum of compound **10** (75 MHz, C$_2$D$_2$Cl$_4$, 298 K).

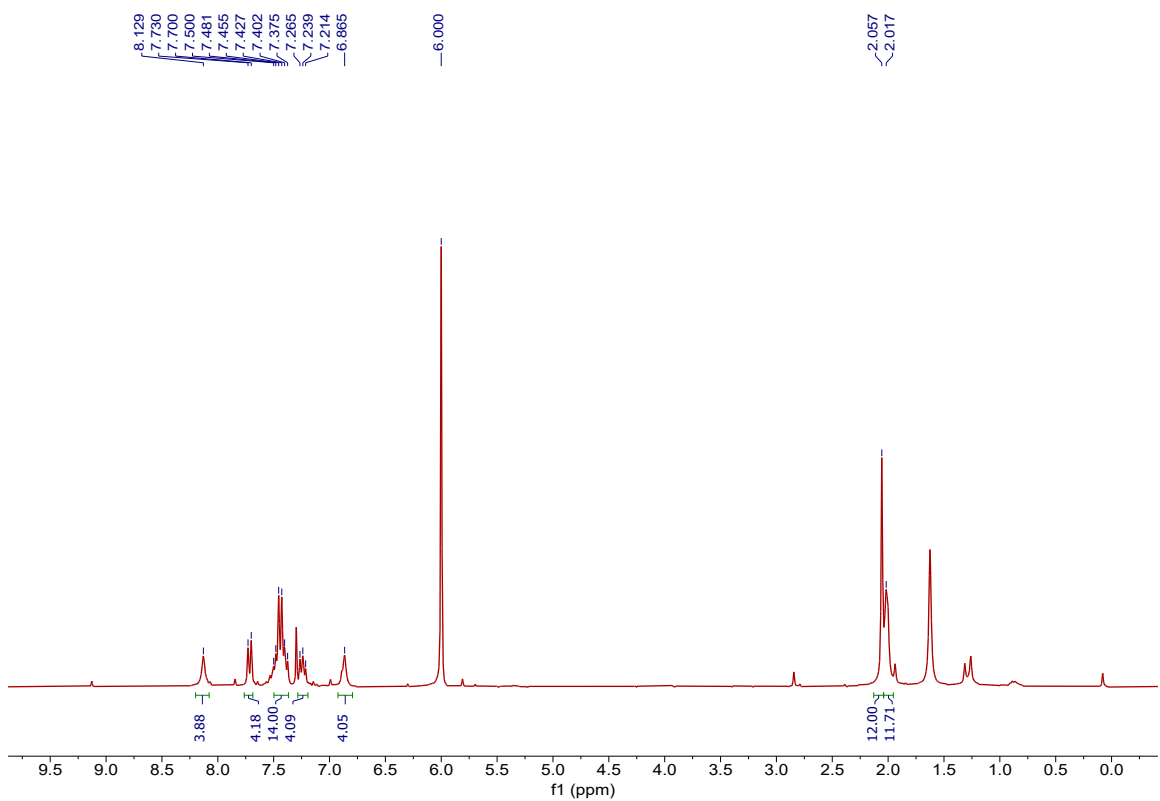

**Supplementary Figure 19** | $^1$H NMR spectrum of precursor **4** (300 MHz, C$_2$D$_2$Cl$_4$, 298 K).

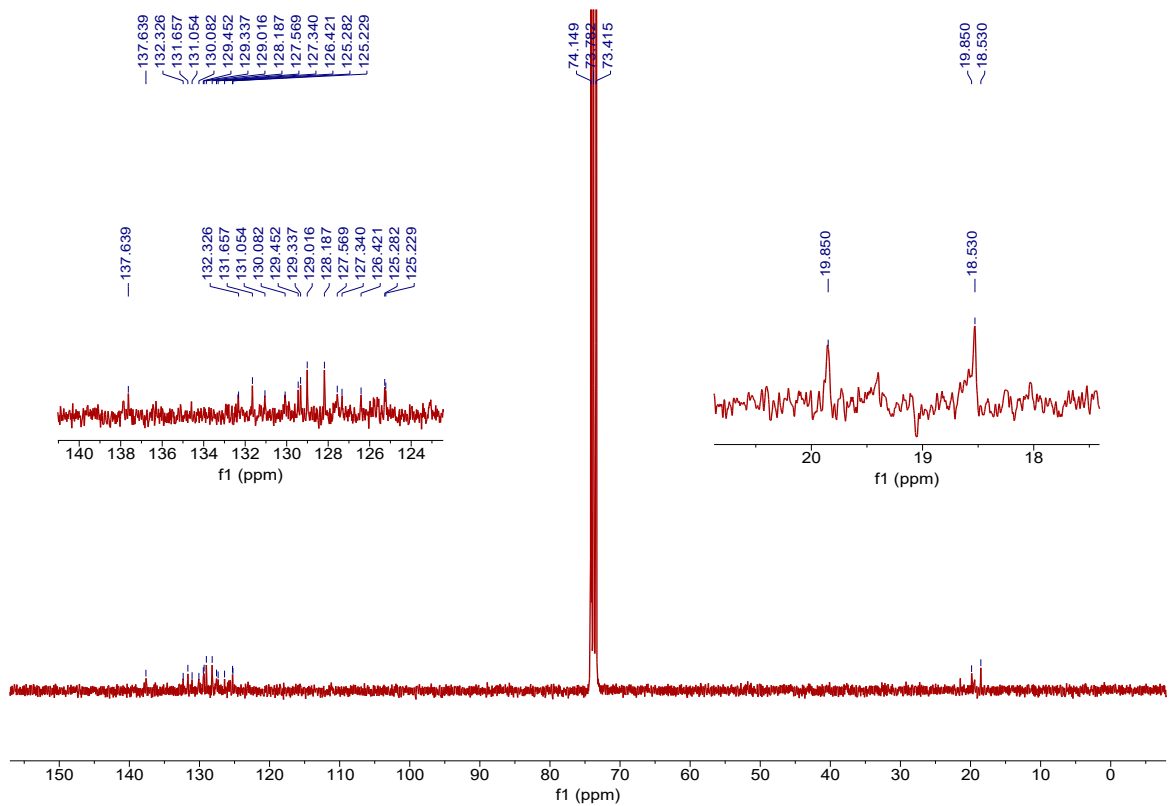

**Supplementary Figure 20** | $^{13}$C NMR spectrum of precursor **4** (75 MHz, C$_2$D$_2$Cl$_4$, 298 K).

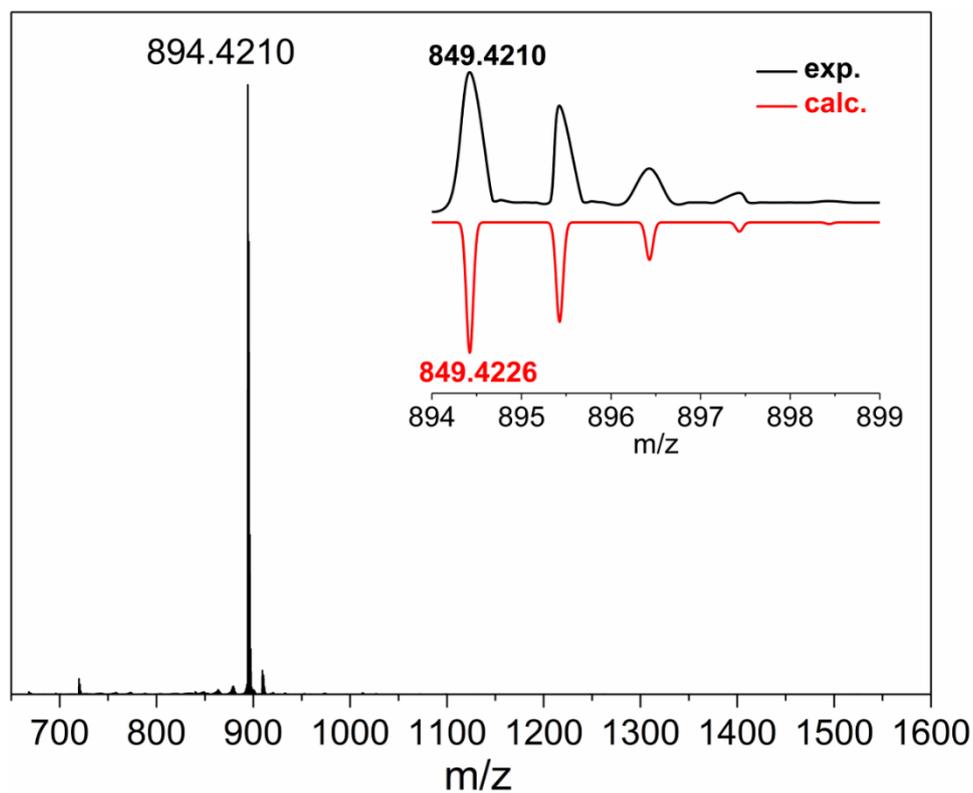

**Supplementary Figure 21 |** High-resolution MALDI-TOF MS spectrum of precursor **4**. Inset displays the isotopic distribution in comparison with the simulated pattern.

## 5. Supplementary References.